\documentclass[fleqn,usenatbib]{mnras}

\usepackage{amsmath,amssymb}
\usepackage{mathptmx}
\usepackage{txfonts}

\usepackage[T1]{fontenc}
\usepackage{ae,aecompl}
\usepackage{graphicx}
\usepackage{amssymb}
\usepackage{bm}
\usepackage{color}
\usepackage{multirow}
\usepackage{booktabs}

\newcommand{\teff}{\ensuremath{T_{\rm eff}}}
\newcommand{\beq}{\begin{equation}}
\newcommand{\eeq}{\end{equation}}
\newcommand{\intd}{{\rm d}}
\newcommand{\msun}{\ensuremath{M_\odot}}

\newcommand{\stefb}{\ensuremath{\sigma_{\rm SB}}}
\newcommand{\mdot}{\ensuremath{\dot{M}}}
\newcommand{\myr}{\ensuremath{\msun\,{\rm yr}^{-1}}}

\newcommand{\ltwo}{L$_2$}
\newcommand{\rltwo}{r_{{\rm L}_2}}
\newcommand{\vesc}{\ensuremath{v_{\rm esc}}}
\newcommand{\vorb}{\ensuremath{v_{\rm orb}}}

\newcommand{\lsun}{\ensuremath{L_\odot}}

\newcommand{\tdiff}{\ensuremath{t_{\rm diff}}}
\newcommand{\tadv}{\ensuremath{t_{\rm adv}}}

\newcommand{\lorb}{\ensuremath{L_{\rm orb}}}

\defcitealias{pmt16}{PMT16}	 	 

\title[Binary Mergers with Marginally-Bound Ejecta]{Binary Stellar Mergers with Marginally-Bound Ejecta: Excretion Disks, Inflated Envelopes, Outflows, and their Luminous Transients}

\author[Pejcha et al.]{ Ond\v{r}ej Pejcha,$^{1}$\thanks{pejcha@astro.princeton.edu}\thanks{Hubble and Lyman Spitzer Jr.\ Fellow}, Brian D.\ Metzger,$^2$ and Kengo Tomida$^3$
\vspace*{6pt}\\
$^1$ Department of Astrophysical Sciences, Princeton University, 4 Ivy Lane, Princeton, NJ 08540, USA\\
$^2$ Columbia Astrophysics Laboratory, Columbia University, New York, NY 10027, USA\\
$^3$ Department of Earth and Space Science, Graduate School of Science, Osaka University, 1-1 Machikaneyama, Toyonaka, Osaka 560-0043, Japan}

\pubyear{2016}

\begin{document}
\maketitle

\begin{abstract}
We study mass loss from the outer Lagrange point (\ltwo) in binary stellar mergers and their luminous transients by means of radiative hydrodynamical simulations. Previously, we showed that for binary mass ratios $0.06 \lesssim q \lesssim 0.8$, synchronous \ltwo\ mass loss results in a radiatively inefficient, dust-forming unbound equatorial outflow.  A similar outflow exists irrespective of $q$ if the ratio of the sound speed to the orbital speed at the injection point is sufficiently large, $\varepsilon \equiv c_T/\vorb \gtrsim 0.15$.  By contrast, for cold \ltwo\ mass-loss ($\varepsilon \lesssim 0.15$) from binaries with $q \lesssim 0.06$ or $q \gtrsim 0.8$, the equatorial outflow instead remains marginally-bound and falls back to the binary over tens to hundreds of binary orbits, where it experiences additional tidal torqueing and shocking.  As the bound gas becomes virialized with the binary, the luminosity of the system increases slowly at approximately constant photosphere radius, causing the temperature to rise.  Subsequent evolution depends on the efficiency of radiative cooling.  If the bound atmosphere is able to cool efficiently, as quantified by radiative diffusion time being shorter than the advection time ($\tdiff/\tadv \ll 1$), then the virialized gas collapses to an excretion disk, while for $\tdiff/\tadv \gtrsim 1$ an isotropic wind is formed.  Between these two extremes, an inflated envelope transports the heat generated near the binary to the surface by meridional flows. In all cases, the radiated luminosity reaches a fraction $\sim 10^{-2}$ to $10^{-1}$ of $\mdot \vorb^2/2$, where $\dot{M}$ is the mass outflow rate.  We discuss the implications of our results for transients in the luminosity gap between classical novae and supernovae, such as V1309~Sco and V838~Mon.
\end{abstract}

\section{Introduction}

The discovery in V1309~Sco of a contact binary with a rapidly decreasing orbital period \citep{tylenda11}, which terminated its evolution in a luminous outburst, established a connection between catastrophic phases of binary star evolution and a class of transients characterized by red colors and luminosities in the gap between classical novae and supernovae \citep{martini99,munari02,bond03,soker03,soker06,tylenda06,kulkarni07,tylenda11,tylenda13,ivanova13b,nandez14,kurtenkov15,smith16}, hereafter collectively denoted as red transients (RT).  Our knowledge of the progenitor binaries of these events has thus far been hindered by the complexity of their photometric and spectral evolution, and the lack of a theoretical framework to interpret these observations.  Obtaining a better understanding of these events is timely because envelope ejection during strong binary interaction may play a crucial role in producing black hole binaries -- the source of recently detected gravitational waves by Advanced LIGO \citep{abbott16a,abbott16b,belczynski16}.  

The durations of RT range from $\sim 20$\,days in the cases of V4332~Sgr and V1309~Sco \citep{martini99,mason10} to $\sim 500$\,days for OGLE-2002-BLG-360 \citep{tylenda13}. Their peak luminosities range from $\sim 10^4$ to $\sim 10^6\,\lsun$, or perhaps even $\gtrsim 10^7\,\lsun$ \citep{smith16}, with effective temperatures ranging from typical values of $\teff \sim 5000$\,K to as low as $800$\,K.  Indeed, much of this diversity can be found within a single RT.  V838~Mon was discovered in outburst in early January 2002 and then remained at nearly constant luminosity of $L\sim 10^5\,\lsun$ and temperature $\teff \sim 5000$\,K for about a month \citep[e.g.][]{tylenda05}. On February 2 of 2002, its luminosity increased by more than an order of magnitude while its temperature simultaneously rose to $\teff \gtrsim 7000$\,K \citep{sobotka02,tylenda05}. Afterwards, V838~Mon evolved to become perhaps the coolest supergiant ever observed, with $\teff \sim 2000$\,K \citep[e.g.][]{evans03}. Similarly complex evolution was seen in V1309~Sco, which exhibited a gradual $\sim 200$\,day-long rise after the periodic variability of the contact binary disappeared \citep{tylenda11}.  This rise time greatly exceeded the binary orbital period of $P\approx 1.44$\,days, suggesting that the merger process was not dynamical, at least initially \citep{pejcha14}.  Slow pre-maximum evolution was observed also in OGLE-2002-BLG-360 \citep{tylenda13}.

RT are common in galaxies like the Milky Way, occuring approximately once every other year \citep{kochanek14}. Although the uncertainties are large, this is already a factor of $\sim 2$ to $3$ times higher than the rate of common envelope events predicted by binary population synthesis \citep{kochanek14}. However, given the diversity of RT, the current sample of events could well be contaminated by stellar collisions \citep[e.g.][]{thompson11,katz12,pejcha13} or other classes of transients unrelated to stellar binaries. 

Binary stars can be driven to strong interaction or merger for several reasons. \citet{rasio92,rasio94,rasio95} and \citet{lai93,lai94a,lai94b,lai94c} investigate the stability of polytropic binary stars of equal mass with respect to secular tidal and dynamical instabilities. After the onset of the secular instability (also known as `Darwin' instability), which is identified as a minimum of angular momentum along an equilibrium sequence, binary evolution proceeds on the synchronization timescale. Dynamical instability drives the stars to merge in a few orbits. The maximum binary separation where these instabilities set in depends on the equation of state (EOS) and the stellar structure, in particular the central concentration \citep[e.g.][]{lombardi11,hwang15}. If the binary overfills its \ltwo\,point before these instabilties are triggered, then the merger is induced by mass and angular momentum loss from \ltwo\ itself, a process which again is expected to occur on the dynamical timescale \citep{lombardi11}. Stable configurations inside the outer Roche lobe are permitted for certain contact binaries with mass ratios $q \ne 1$ \citep{rasio95_wuma,li06}.

If one of the stars fills its Roche lobe before the other, the ensuing dynamically unstable mass transfer can also drive the binary together. The precise condition for the initiation and behavior of this runaway depends on the non-adiabatic response of the surface layers of the mass transferring star \citep[e.g.][]{hjellming87,ge10,ge15,passy12,pavlovskii15}. A similar sensitivity to surface layer physics is expected in mergers driven by \ltwo\ mass loss, but less so for those instigated by secular and dynamical instabilities. We note that \ltwo\ spiral streams accompany even dynamical instability \citep[e.g.][]{rasio95,lombardi11} and that mass can be unbound through the \ltwo/L$_3$ points even in the case of L$_1$ mass transfer \citep[e.g.][]{sytov07,sytov09}.

The pre-maximum light curve of V1309~Sco is best understood by \ltwo\ mass loss lasting for thousands of orbital periods \citep{pejcha14}, much longer than the time from contact to coalescence predicted by numerical simulations \citep{nandez14}. This observation motivates exploring the dynamics and observational appearance of quasi-steady \ltwo\ mass loss which occurs over many orbits.  

Indeed, well before V1309~Sco, \citet{kuiper41} first investigated the structure of \ltwo\ outflows, leading to a series of analytic works culminating in \citet{shu79}.  In \citet[hereafter PMT16]{pmt16}, we performed the first radiation hydrodynamics simulations of long-lived \ltwo\ mass loss.  We showed that gas launched synchronously from \ltwo\ is unbound due to tidal torquing for binary mass ratios of $0.064 \lesssim q \lesssim 0.78$, in agreement with the analytic result of \citet{shu79}.  \citetalias{pmt16} also showed that the spiral stream structure merges at a radial distance of roughly 10 times the semi-major axis, through a radial shock which thermalizes $\sim 5\%$ of the outflow kinetic energy.  This shocked ejecta radiates with a luminosity and effective temperature which depends on the binary parameters and mass loss rate. Broadly speaking, the outflow is radiatively inefficient and, because the gas cools to low temperatures through expansion and radiation, dust forms in copious amounts.  \citetalias{pmt16} predicted a correlation between the expansion velocity and the transient luminosity, which is roughly obeyed by the known RT.  The appearance of the transient is strongly dependent on viewing angle because the optical depth of the equatorially-focused outflow is lower along the vertical direction parallel to the binary axis. 

This paper extends the analysis of \citetalias{pmt16} to binaries with $q \lesssim 0.064$ or $q \gtrsim 0.78$, for which analytic models of \citet{shu79} instead predict inefficient tidal torqueing and the formation of a bound circumbinary excretion disk. We also relax the assumption of negligible thermal content of the material at \ltwo. In Section~\ref{sec:setup}, we briefly describe the setup of our calculations and emphasize the differences with respect to \citetalias{pmt16}. In Section~\ref{sec:results}, we show that the excretion disk is formed only when radiative cooling is efficient, and characterize other possible outcomes: an isotropic or equatorial wind and an inflated envelope. We also discuss the luminosity and effective temperature evolution and the backreaction on the central binary. In Section~\ref{sec:summary}, we give an overview of the \ltwo\ mass loss outcomes, effectively summarizing the results of this paper and \citetalias{pmt16}. In Section~\ref{sec:implications}, we conclude by discussing the implications for the red transients.

\section{Calculation setup}
\label{sec:setup}

Our simulation set up closely follows \citetalias{pmt16}, to which we refer the reader for additional details not described here.  We employ smoothed particle hydrodynamics (SPH) with variable smoothing lengths \citep{price07}.  The acceleration of each particles is calculated as
\beq
\frac{\intd \mathbf{v}}{\intd t} = \mathbf{a}_{\rm hydro} + \mathbf{a}_{\rm visc} + \mathbf{a}_{\rm binary},
\label{eq:acceleration}
\eeq
where $\mathbf{a}_{\rm hydro}$ and $\mathbf{a}_{\rm visc}$ are the standard gas and viscous forces \citep{monaghan83,balsara95}, respectively, and $\mathbf{a}_{\rm binary}$ is the acceleration from the combined gravitational potential of the central binary. The binary is modeled as two point masses $M_1$ and $M_2$, $M_1<M_2$, on a circular Keplerian orbit in the $x-y$ plane with semi-major axis $a$, orbital period $P$, and the center of the mass as the coordinate system origin. The binary parameters do not change in time, because our goal is to first understand the hydrodynamics of the circumbinary gas in the quasi-stationary limit.  Gravitational attraction between particles is neglected, but gravitational clumping is not expected to be important \citepalias{pmt16}. 

The specific internal energy of individual particles, $u$, evolves as
\beq
\frac{\intd u}{\intd t} = \dot{u}_{\rm hydro} + \dot{u}_{\rm visc} + \dot{u}_{\rm diff} + \dot{u}_{\rm cool},
\label{eq:energy}
\eeq
where $\dot{u}_{\rm hydro}$ and $\dot{u}_{\rm visc}$ are standard SPH terms describing adiabatic expansion or contraction and viscuous heating. Energy is redistributed between the particles assuming flux-limited diffusion $\dot{u}_{\rm diff}$ \citep[e.g.][]{bodenheimer90,forgan09}, and particles are allowed to radiatively cool through the term $\dot{u}_{\rm cool}$ \citep{stamatellos07,forgan09}. Irradiation by the central binary is neglected, because it only slightly modifies the temperature structure of the stream very close to the binary and does not have an effect on the stream properties further out. More importantly, the outflow luminosities we find here are much higher that the luminosity of the central binary star and the irradiation calculation noticably slows down the computation.

Similarly to \citetalias{pmt16}, we use the solar-metallicity equation of state of \citet{tomida13,tomida15}, which takes into account ionization of hydrogen and helium and molecular states of H$_2$. In this work, we use the opacity tables compiled by \citet{tomida13}, which are based on \citet{semenov03}, \citet{ferguson05}, and the Opacity Project \citep{seaton94}. Opacities outside of the coverage of the original tables are extrapolated based on the boundary values. With the updated opacity tables, we are able to distinguish between the Rosseland and Planck means in the prescription for radiative cooling
\beq
\dot{u}_{\rm cool} = -\frac{\stefb T^4}{\Sigma_z\tau_z + \kappa_{\rm P}^{-1}}
\eeq
where $\Sigma_z$ and $\tau_z$ are the surface density and optical depth, respectively, from the position of the particle outwards in the direction perpendicular to the equatorial plane, and $\kappa_{\rm P}$ is the Planck-mean opacity. Both $\tau_z$ and $\dot{u}_{\rm diff}$ are calculated using the Rosseland mean opacities.

As in \citetalias{pmt16}, we calculate the radiated luminosity $L$ as the sum of radiative cooling of all particles
\beq
L = \sum_i m_i \dot{u}_{\rm cool},
\eeq
where $m_i$ is the mass of particle $i$. The effective temperature $\teff$ is estimated as the radiative cooling-weighted mean of effective temperatures of individual particles
\beq
\teff^4 = \frac{1}{L}\sum_i m_i \dot{u}_{\rm cool} \frac{T_i^4}{\tau_z + 1}.
\eeq

Following \citetalias{pmt16}, we inject the particles in a region of size $\varepsilon a$ near the \ltwo\ point, where $\varepsilon \equiv c_T/\vorb< 1$ is the ratio of the gas sound speed $c_T$ to binary orbital velocity $\vorb = \sqrt{GM/a}$. The radial position of the \ltwo\ point $\rltwo$ is obtained by usual means \citep{shu79}. In most of our runs, the injected particles possess a constant temperature equal to the surface temperature of the binary $T_{\rm binary} = 4500$\,K and we adopt a fixed value of $\varepsilon = 0.05$.  However, we also explore a limited number of models with a higher value of $\varepsilon$, corresponding to considerably hotter ejecta.  The latter aims to capture situations in which the relatively cool thin surface layers of the binary are stripped rapidly after the onset of \ltwo\ mass loss, exposing hotter layers of the star which have not had enough time to radiatively cool.  We offset the injection point outward by $10^{-3}a$ along the axis connecting the two stars to reduce the number of particles that are immediately re-absorbed by the binary. Particles are injected with a constant mass $\mdot/\dot{N}$, where $\mdot$ is the mass increase rate of active particles in the simulation and the number injection rate $\dot{N}$ specifies the resolution of the simulation, which we typically take to be $\dot{N} = 1000/P$. The evolution is followed for $\gtrsim 100$ binary orbital periods, with each run typically consuming three to seven days on a 20-core machine. The main limitation is the explicit timestep in the latest stages of the evolution, when radiative processes are important.

The inner boundary condition requires special care, as it could be important for marginally-bound outflows that fall back to the vicinity of the binary. As in \citetalias{pmt16}, we employ a default outflow inner boundary condition, where we simply remove particles that fall within the radius $\rltwo$ of the binary barycenter. Our definition of $\mdot$ implies that the particles removed at the inner boundary are compensated for by the injection of new particles at \ltwo\ to maintain the prescribed growth of total mass of active particles. This assumption that absorbed particles are immediately re-emitted from \ltwo\ is necessarily a simplification.

Additionally, we implement a reflecting inner boundary condition in the form a sphere with a radius $\rltwo$ around the barycenter. We construct ghost particles by mirroring the particles that approach the reflecting boundary according to the prescription of \citet{herant94}. The ghost particles share most properties with their active counterparts, with only the normal component of their velocity being inverted; this is equivalent to a free slip along the boundary \citep[e.g.][]{libersky93,colagrossi03}. We tested the reflecting boundary on a test problem of a thin cold ideal-gas shell with a net angular momentum free-falling in a gravitational potential of a point mass. We find good conservation of energy and angular momentum during the dynamic stages, but the conservation breaks during the subsequent viscous evolution of the rotating bound atmosphere.  Although the details depend on precisely how the ghost particle properties are assigned, we have not been able to obtain fully conservative long-term viscous evolution. This is, however, not surprising since we are not properly taking into account the boundary layer of the central object.  Likewise, in the physical problem of interest the inner boundary is not simply a sphere, but rather two distorted stars, which can mechanically shock/stir or reabsorb matter which falls back to the binary. Despite these complications, we find that the overall picture of the hydrodynamic evolution does not depend sensitively on the inner boundary condition.

\begin{table*}
\caption{List of simulations. For each run characterized by semi-major axis $a$, orbital period $P$, stellar masses $M_1$ and $M_2$, and the mass-loss rate $\mdot$, we show the duration of the simulation $t_{\rm max}$, luminosity at the end of the simulation $L_{\rm final}$, and the classification of the outcome as an isotropic wind (IW), equatorial wind (EW), inflated envelope (IE), and circumbinary disk (CBD). The default parameters of the simulations were $T_*=4500$\,K, $\varepsilon=0.05$, $\dot{N}=1000/P$, and inflow/outflow inner boundary condition. The remarks describe modifications with respect to these default values. Reflective inner boundary condition is abbreviated as RIB.}
\label{tab:list}
\begin{tabular}{ccccccccl}
\hline
$a$ & $M_1$ &  $M_2$ & $P$ & $\mdot$ & $t_{\rm max}$ & $L_{\rm final}$ & Outcome & Remark\\
AU  & $\msun$ & $\msun$ & days & $\myr$ &  $P$ & $10^3 \lsun$ & & \\
\hline
\multirow{3}[5]{*}{0.0045} & \multirow{3}[5]{*}{0.8} & \multirow{3}[5]{*}{0.9} & \multirow{3}[5]{*}{0.085} & $10^{-1}$ &  121.5 & 28 &IW &\\\cmidrule{5-9}
 &  &  &  & $10^{-2}$&  118.8 & 33 & IW & \\\cmidrule{5-9}
 &  &  &  & $10^{-3}$ &  125.6 & 14 & CBD/IE &\\\cmidrule{1-9}
\multirow{16}{*}{0.03} & \multirow{16}{*}{0.8} & \multirow{16}{*}{0.9} & \multirow{16}{*}{1.45} & \multirow{5}{*}{$10^{-1}$} &  125.4 & 28 & IW &\\
&  &  &  & & 114.4 & 25 & IW & $\dot{N}=500/P$\\
 &  &  &  & &  96.6 & 21 & IW & $\dot{N}=4000/P$\\
&  &  &  & & 112.9 & 38 & IW & RIB\\
&  &  &  & &  100.4 & 17 & EW & $T_*=40000$\,K, $\varepsilon=0.15$\\\cmidrule{5-9}
&  &  & & \multirow{4}{*}{$10^{-2}$} & 110.8 & 11 & IW/IE & \\
&  &  &  & & 112.9 & 14 & IW & RIB\\
&  &  &  & & 143.5 &  & IW & no cooling, no diffusion\\
&  &  &  & & 123.0&  & IW & no cooling, no diffusion, RIB\\\cmidrule{5-9}
&  &  & & \multirow{6}{*}{$10^{-3}$} & 127.2 & 2.0 & CBD & \\
&  &  &  & & 124.0 & 2.8 & CBD & RIB\\
&  &  &  & & 108.3& 2.0 & CBD& $T_*=40000$\,K, $\varepsilon=0.15$\\
&  &  &  & & 53.5& 2.0 & EW & $T_*=100000$\,K, $\varepsilon=0.15$\\
&  &  &  & & 288.5 &  & IW & no cooling, no diffusion\\
&  &  &  & & 196.1&  & IW & no cooling, no diffusion, RIB\\\cmidrule{5-9}
&  &  & & $10^{-4}$ & 82.0 & 0.22 & CBD & \\\cmidrule{1-9}
\multirow{5}[8]{*}{0.03} & \multirow{5}[8]{*}{0.1} & \multirow{5}[8]{*}{2.0} & \multirow{5}[8]{*}{1.31}  & $10^{-1}$ &  106.1 & 28 & IW & \\\cmidrule{5-9}
 &  &  & & \multirow{2}{*}{$10^{-2}$} &  143.2 & 14 & IE/IW\\
&  &  &  & &  102.9 & 20 & IW & RIB\\\cmidrule{5-9}
&  &  & & $10^{-3}$ &  124.2 & 2.2 & CBD &\\\cmidrule{5-9}
&  &  & &  $10^{-4}$ &  79.8 & 0.20 & CBD &\\\cmidrule{1-9}
\multirow{4}[7]{*}{0.2} & \multirow{4}[7]{*}{0.8} & \multirow{4}[7]{*}{0.9} & \multirow{4}[7]{*}{25.1} & 1.0 &  136.6 & 54 & IW & \\\cmidrule{5-9}
&  &  & &$10^{-1}$ &  145.0 & 13.6 & IE &\\\cmidrule{5-9}
&  &  & &$10^{-2}$ &  117.6 & 2.8 & CBD & \\\cmidrule{5-9}
&  &  & &$10^{-3}$ &  89.8 & 0.3 & CBD\\\cmidrule{1-9}
\multirow{5}[8]{*}{0.2} & \multirow{5}[8]{*}{4.0} & \multirow{5}[8]{*}{4.5} &\multirow{5}[8]{*}{11.2} & \multirow{2}{*}{$1.0$} &  117.5 & 352 & IW\\
 &  &  & &  &  112.3 & 429 & IW & RIB\\\cmidrule{5-9}
 &  &  & &\multirow{2}{*}{$10^{-1}$} &  133.8 & 70 & IE/CBD &\\
&  &  & & &  111.8 & 102 & IE/CBD & RIB\\\cmidrule{5-9}
&  &  & &$10^{-2}$ &  90.1 & 15 & CBD &\\\cmidrule{5-9}
&  &  & &$10^{-3}$ &  75.9 & 2.0 & CBD &\\\cmidrule{1-9}
\multirow{5}[8]{*}{1.3} & \multirow{5}[8]{*}{0.8} & \multirow{5}[8]{*}{0.9} &\multirow{5}[8]{*}{415} & $10$ &  131.1 & 66 & EW &\\\cmidrule{5-9}
 &  &  & &$1.0$ &  80.0 & 24 & EW & \\\cmidrule{5-9}
 &  &  & &\multirow{2}{*}{$10^{-1}$} &  71.4 & 8.0 & EW &\\
&  &  & & &  23.4 & 5.0 & EW & polytropic EOS with $\Gamma =5/3$\\\cmidrule{5-9}
&  &  & & $10^{-2}$ &  91.8 & 0.6 & CBD & $T_* = 3000$\,K\\
\hline
\end{tabular}
\end{table*}

In Table~\ref{tab:list}, we give the summary of our simulations. Our goal is to understand the hydrodynamics and radiative properties of the ultimate steady-state of the binary mass-loss. We thus do not self-consistently evolve $\mdot$ and the parameters of the binary orbit. As as result, some of the runs shown in Table~\ref{tab:list} are not entirely realistic. For example, for runs with large $a$ the total mass lost is comparable to the mass of the binary, but we do not include self-gravity of the ejecta and do not change the orbital parameters. Such runs serve primarily to verify our analytic estimates over wider range of parameter space.

\section{Results}
\label{sec:results}

\subsection{Marginally-bound outflows}
\label{sec:bound}

\begin{figure*}
\centering
\includegraphics[width=0.49\textwidth]{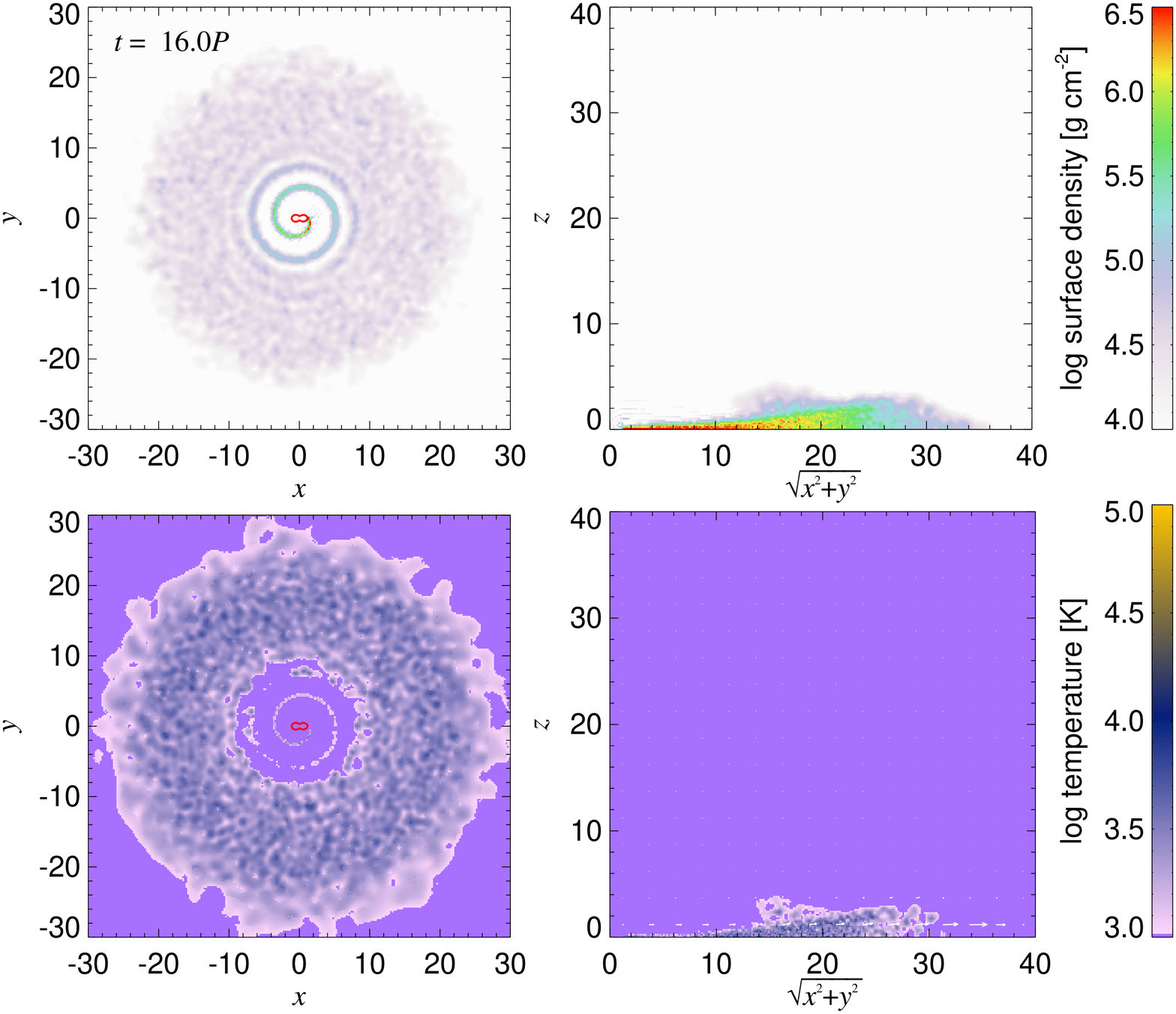}\hfill\includegraphics[width=0.49\textwidth]{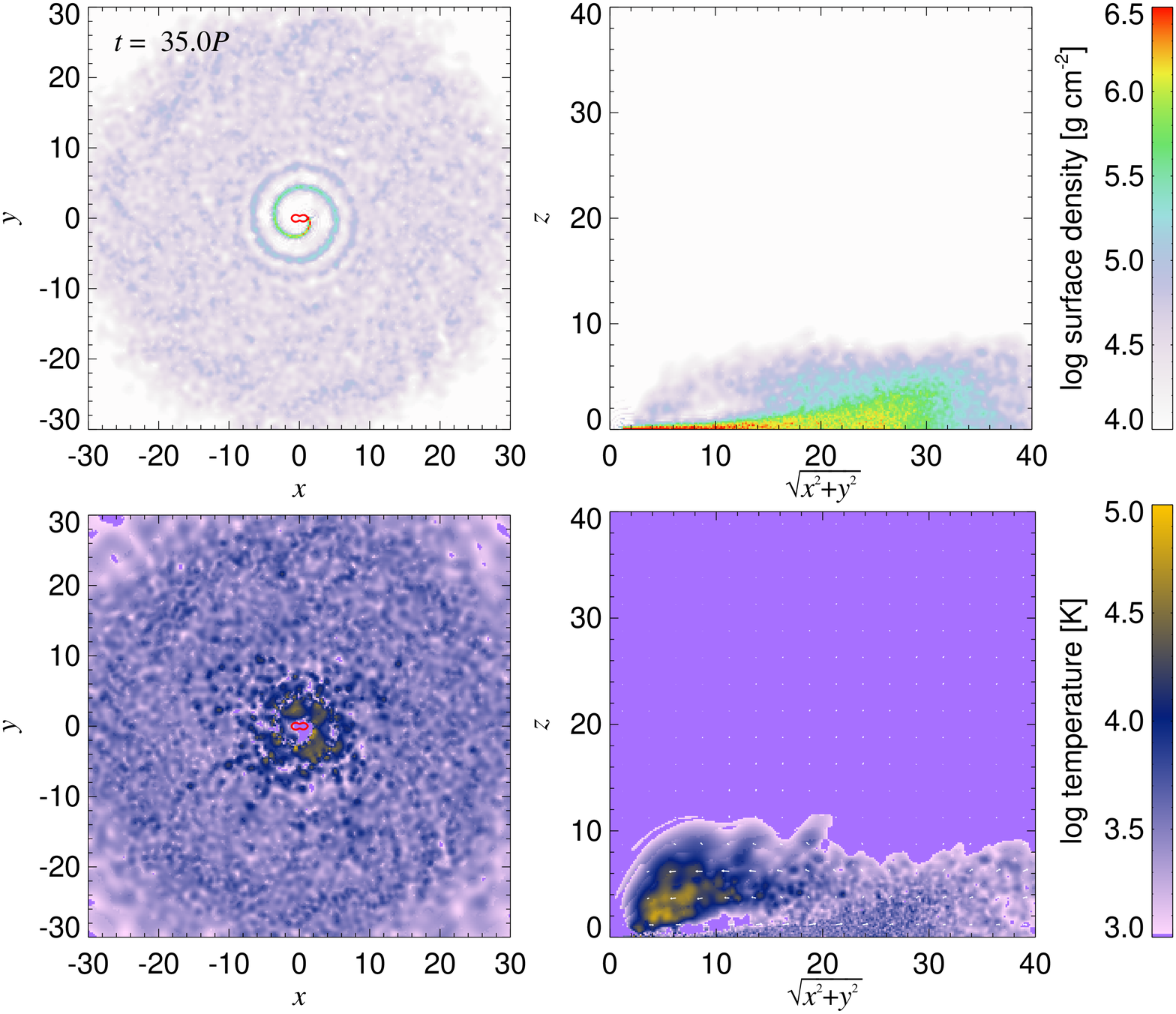}

\vspace{0.3cm}

\includegraphics[width=0.49\textwidth]{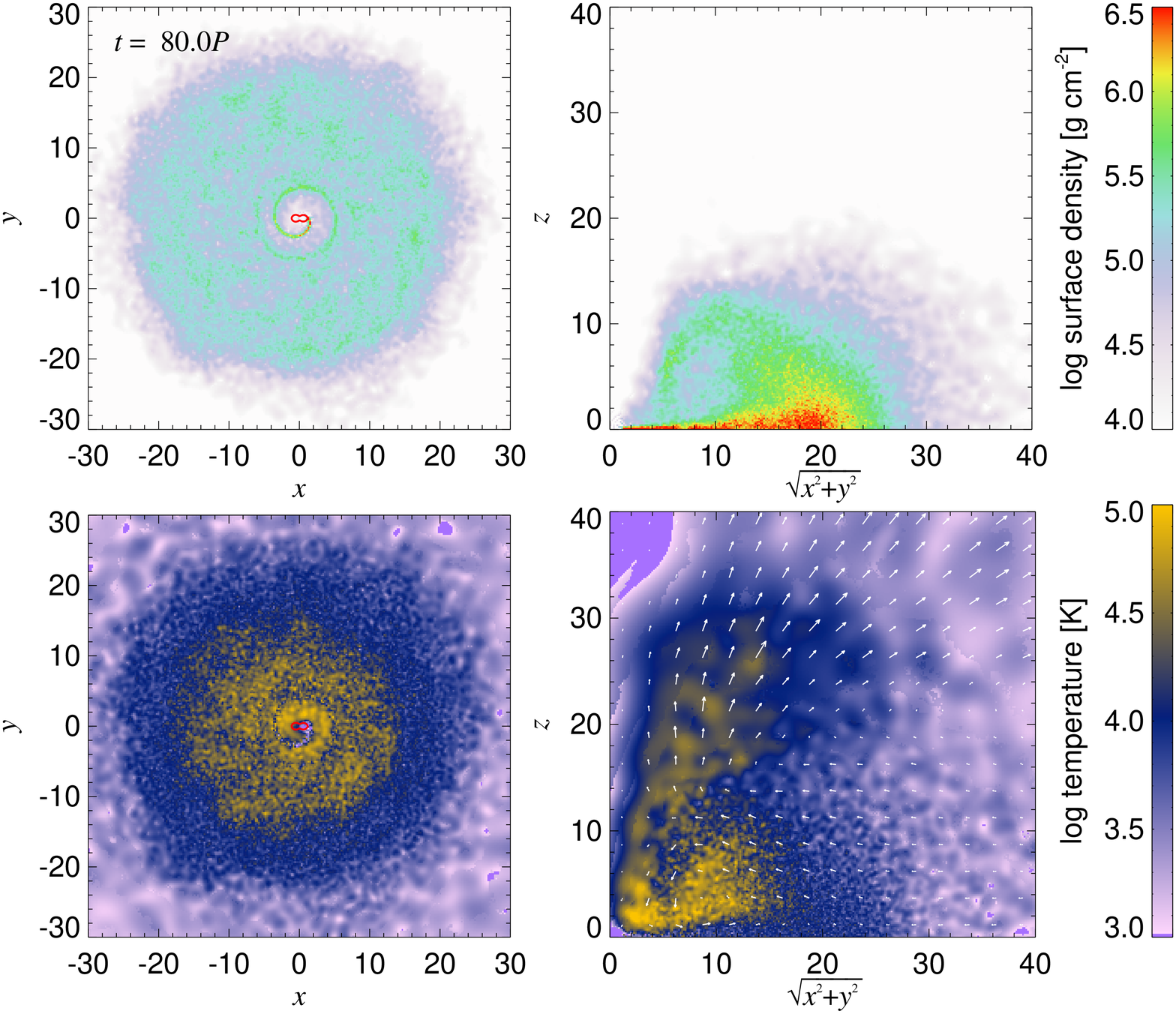}\hfill\includegraphics[width=0.49\textwidth]{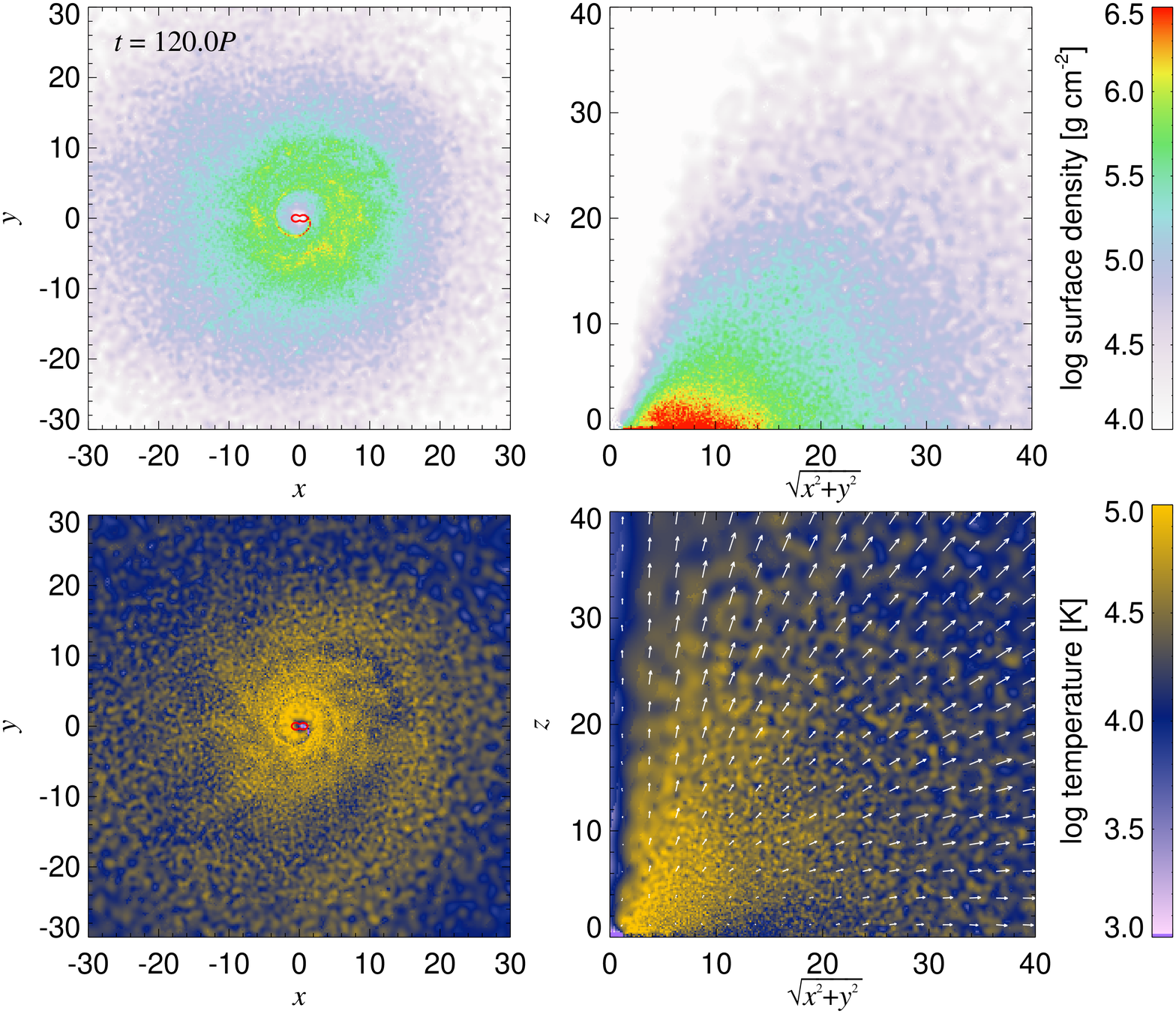}
\caption{\label{fig:anim_all} Stages in the evolution of a marginally-bound \ltwo\ outflow with mass loss rate $\mdot=0.1\,\myr$ from a binary with $a=0.03$\,AU, $M_1=0.8\,\msun$, and $M_2=0.9\,\msun$. The four sets of panels show snapshots at epochs $t/P = 16$ (top left), $35$ (top right), $80$ (bottom left), and $120$ (bottom right) after the initiation of the mass loss.  Each of the four sets of panels shows surface density and temperature structure of the outflow visualized in the inertial $x-y$ plane as well as in the cylindrical coordinates $\sqrt{x^2+y^2}-z$. White arrows show the velocity field with scale such that the distance between two velocity vector origins corresponds to $\vesc/3$. Spatial coordinates are in the units of binary semi-major axis $a$.}
\end{figure*}

\begin{figure*}
\centering
\includegraphics[width=0.49\textwidth]{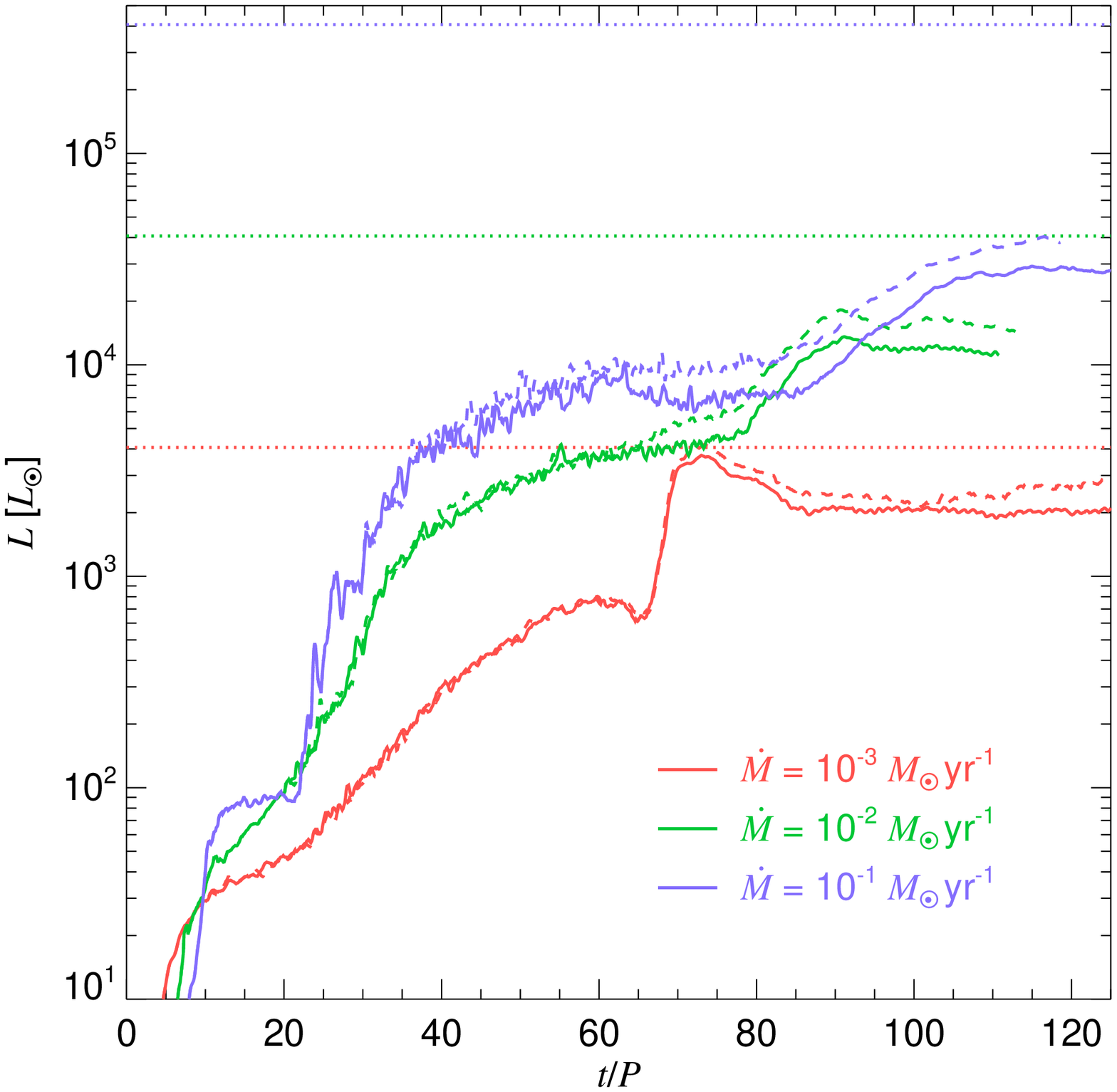}
\includegraphics[width=0.49\textwidth]{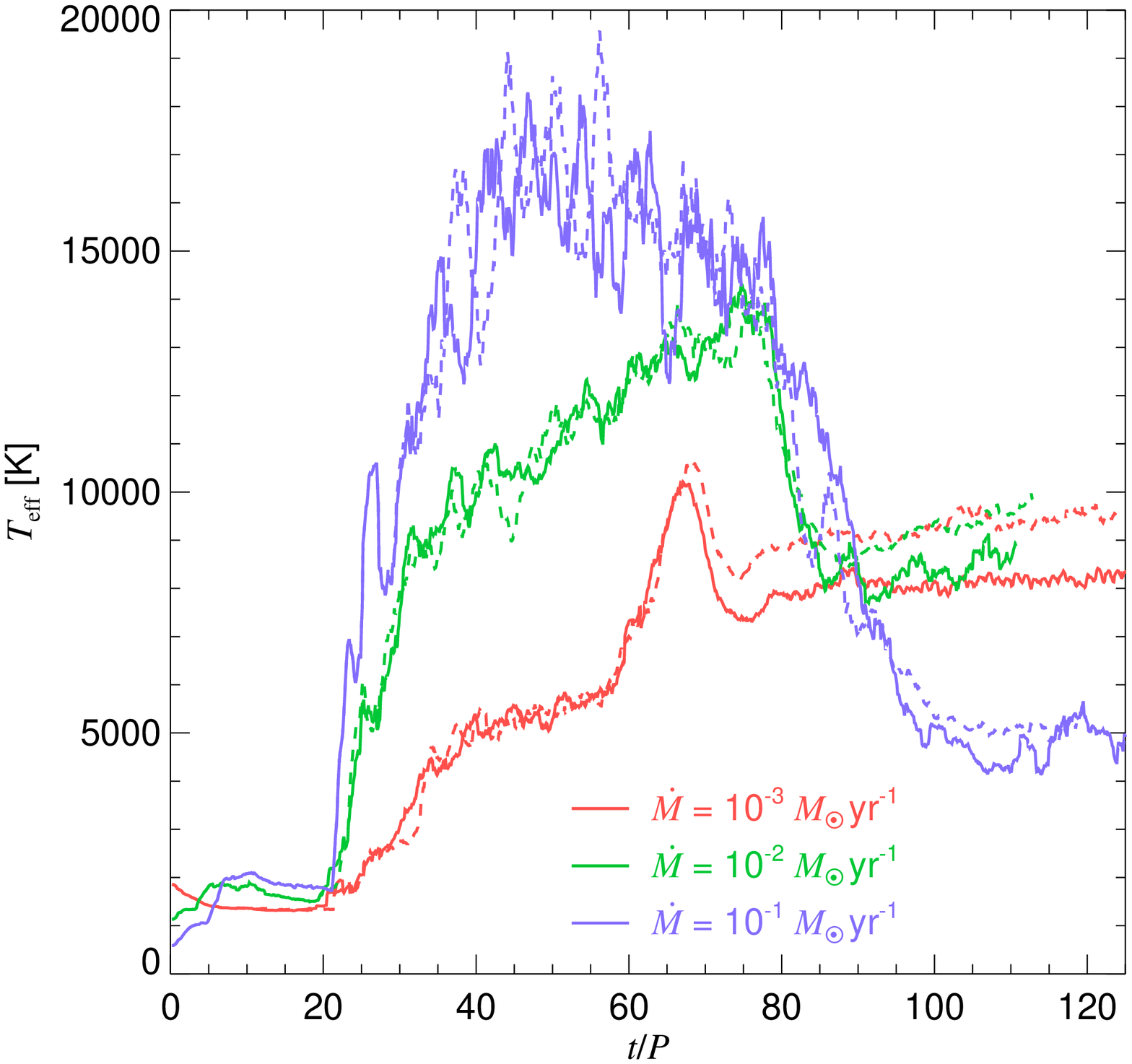}
\caption{\label{fig:lc} Luminosity and effective temperature evolution for a binary with $a=0.03$\,AU, $M_1=0.8\,\msun$, $M_2=0.9\,\msun$, and three different values of $\mdot$ indicated in the legend. Solid and dashed lines show simulations which employ outflow and reflecting inner boundary condition, respectively. Horizontal dotted lines show the maximum achievable luminosity $L_{\rm orb}$ (Eq.~[\ref{eq:lorb}]).}
\end{figure*}

Figure~\ref{fig:anim_all} shows the density and temperature structure of the outflow at representative times illustrating different phases of the evolution. Figure~\ref{fig:lc} shows the corresponding radiative luminosity and an estimate of the effective temperature\footnote{Movies of these figures are available in the online version and at \url{http://www.astro.princeton.edu/~pejcha/ltwo}.}. During the initial phase lasting $\sim 20P$, the dynamics follows the evolution described previously in \citetalias{pmt16}. Specifically, the gas leaves the \ltwo\ point in a spiral, which is tidally torqued by the central binary. The spiral arms merge and thermalize a small fraction of the kinetic energy, which leads to a luminosity of $\sim 100\,\lsun$ with $\teff \sim 1500$\,K. However, a small fraction of the gas has already stalled and starts falling back to the binary, as can be seen at cylindrical radius $\sqrt{x^2+y^2} \approx 15a$ and $z \approx 2a$ in the top left set of panels of Figure~\ref{fig:anim_all}.

Gradually, more material returns to the binary above and below the equatorial outflow, as can be seen in the top right set of panels of Figure~\ref{fig:anim_all}. The returning matter typically has enough angular momentum to avoid absorption or reflection by the inner boundary. Instead, the gas scatters in the time-changing gravitational field of the binary and leaves again in nearly random direction close to the orbital plane. As a result, the gas is shock-heated to nearly virial temperatures of the binary orbit. As more material returns to the binary, the luminosity and effective temperature slowly rise. The increase in $L$ is primarily driven by the increase in $\teff$, because the outer edge of the ejecta is slowly receding from the maximum radius of $\sim 25a$.  Note that our estimate of $\teff$ assumes that system is viewed face-on, but other inclinations would likely yield lower $\teff$. The duration of the slow brightening phase is approximately the time it takes a ballistic particle to fly to $\sim 25a$ and back, which is typically $\sim 100P$.  This introduces into the emission evolution a timescale considerably longer than the orbital period.

When the bulk of the ejecta falls back to the binary (lower left set of panels in Fig.~\ref{fig:anim_all}), the dense cold spiral cannot penetrate through the outer rim of the ejecta and its kinetic energy is thermalized. Simultaneously, more material is heated in the vicinity of the binary, a fraction of which becomes unbound. The dense equatorial belt restricts the outflow in the direction perpendicular to the orbital plane. However, only a very small amount of material is ejected in this phase, as is visible in the panel of vertical density structure. At this stage, the luminosity and effective temperature asymptote to nearly constant values.

In the final stage of the outflow evolution, most of the ejecta has returned to the binary.  The \ltwo\ spiral stream disrupts at a radius of $\sim 4a$, resulting in prolonged heating of the gas by spiral shocks induced by the binary motion. The weak polar outflow expands to fill all solid angles, resulting in an isotropic outflow with a mass loss rate which nearly equals the mass injection rate from \ltwo. The wind is driven thermally from a region near the binary, where the radial velocity nearly vanishes.  The luminosity increases by a factor of $\sim 10$ and $\teff$ decreases to $\approx 5000\,K$, which reflects the fact that the wind is more radiatively efficient due to larger surface area of the photosphere, which is positioned near the hydrogen recombination front.

\begin{figure}
 \centering
\includegraphics[width=\columnwidth]{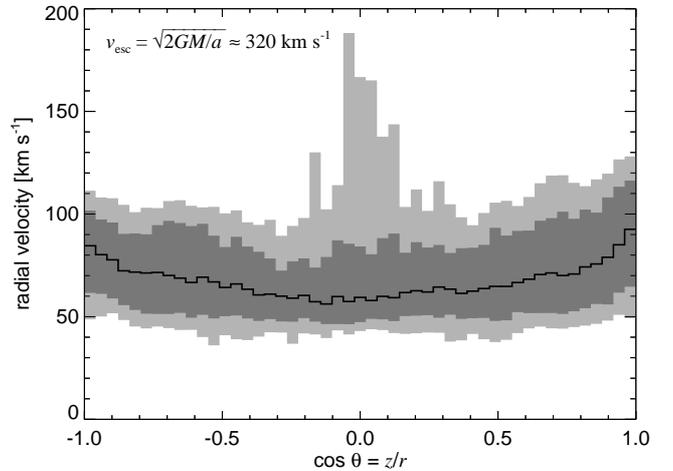}
\caption{\label{fig:vel_theta} Distribution of wind terminal velocities as a function of inclination $\cos \theta = z/r$ for the simulation shown in Figure~\ref{fig:anim_all}.  The region shaded light (dark) grey includes the central $98\%(80\%)$ of the particles, while the black line shows the median.}
\end{figure}

The outflow is nearly isotropic, with slightly higher velocities along the poles than in the orbital plane, as shown in Figure~\ref{fig:vel_theta}. The typical terminal velocity is $\approx 0.25\vesc$, where $\vesc = \sqrt{2GM/a}$ is the binary escape speed.  A small fraction of the gas is ejected with higher velocities along the orbital plane at early times in the simulation.

\subsection{Wind, convective envelope, excretion disk}

\begin{figure}
 \centering
\includegraphics[width=0.7\columnwidth]{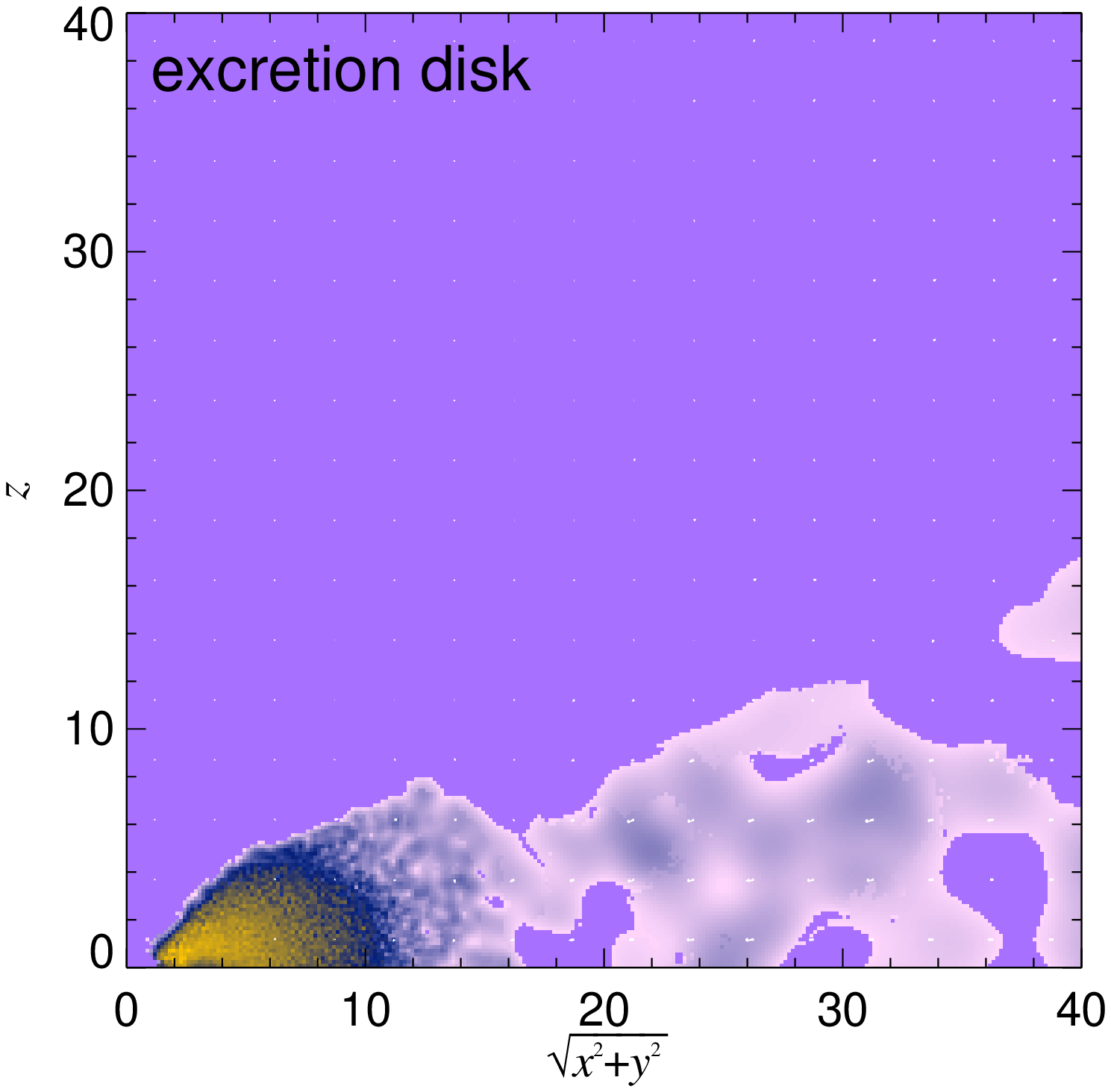}
\includegraphics[width=0.7\columnwidth]{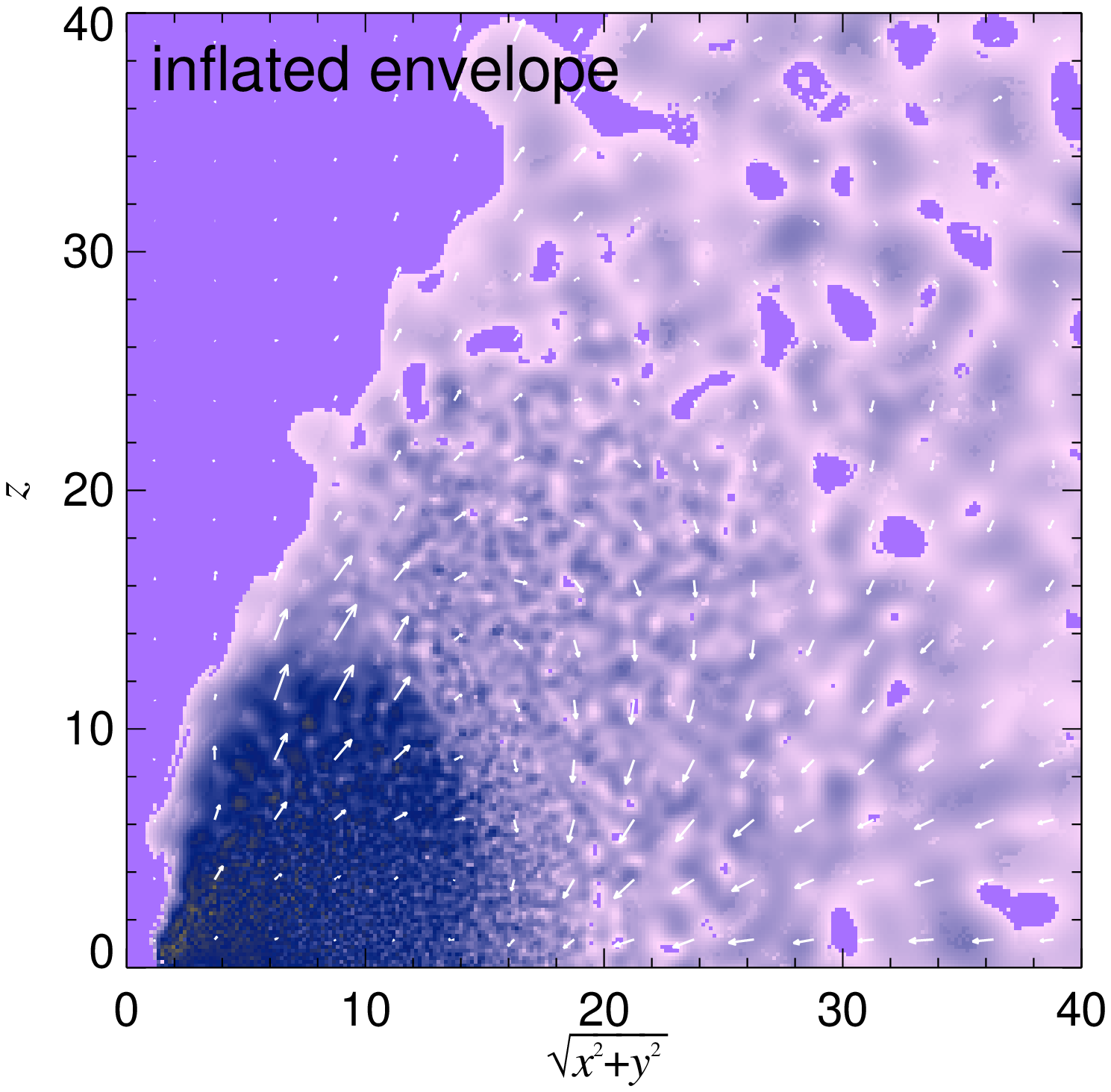}
\caption{\label{fig:f} Two additional final steady-state configurations of the marginally-bound \ltwo\ outflow illustrated by the vertical temperature structure and velocity field.  Top panel shows a rotationally-supported disk at $t=125P$ for the same binary as in Figure~\ref{fig:anim_all} but with $\mdot = 10^{-3}\,\myr$. The bottom panel shows an inflated convective envelope at $t=135P$ for a binary with $a=0.2$\,AU, $M_1=0.8\,\msun$, $M_2=0.9\,\msun$, and $\mdot = 10^{-1}\,\myr$.  The color coding and symbol labels have the same meaning as in Figure~\ref{fig:anim_all}, except the bottom panel, where the velocity vector scale is such that the distance between two velocity vector origins corresponds to $\vesc/6$.}
\end{figure}

The evolution of the system depends qualitatively on the mass-loss rate and the binary semi-major axis.  Figure~\ref{fig:f} shows the final configuration of the marginally-bound \ltwo\ outflow for two additional models. The top panel shows a rotationally supported disk, while the bottom panel shows an inflated envelope. In the latter case, a meridional flow transports the energy generated near the binary outward, where it can be efficiently radiated, after which point the gas returns to the binary to repeat the cycle. No continuous outflow is achieved in either case. 

Table~\ref{tab:list} summarizes the outcomes of our simulations. The primary means for the classification was the morphology and vertical velocity structure of the ejecta near the end of the runs as well as the behavior of total unbound mass and specific angular momentum. Recognizing isotropic wind and circumbinary disk is relatively straightforward, but we observed only one clear case, where the final state was the inflated envelope (shown in Fig.~\ref{fig:f}). In a few cases, the final morphology had aspects of both inflated envelope and circumbinary disk or isotropic outflow. We also indicate such ambiguity in Table~\ref{tab:list}.

\begin{figure*}
 \centering
\includegraphics[width=0.8\textwidth]{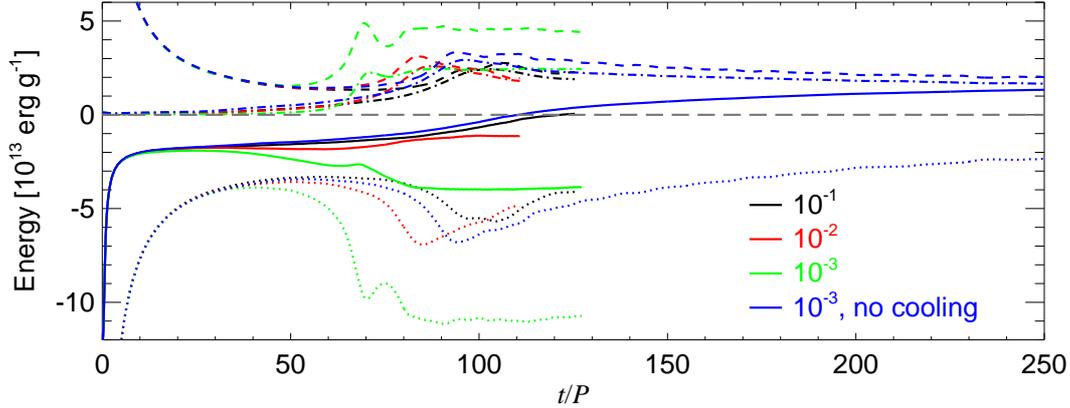}
\caption{\label{fig:energy} Time evolution of the average specific total (solid lines), potential (dotted), kinetic (dashed), and thermal (dash-dotted) energies for a binary with $a=0.03$\,AU, $M_1 = 0.8\,\msun$, $M_2=0.9\,\msun$, and three values of $\mdot$ indicated in the units of $\myr$ in the legend. Blue lines show an identical calculation for $\mdot=10^{-3}\,\myr$, but with radiative diffusion and cooling artificially neglected to illustrate that the radiative cooling efficiency is the determining factor for the final configuration. The sum of the potential and kinetic energies at the \ltwo\ point is approximately given by $-1.4\times 10^{14}$\,ergs\,g$^{-1}$.}
\end{figure*}

To illustrate the differences between the final configurations quantitatively, Figure~\ref{fig:energy} shows the time evolution of the specific energy for a range of simulations with different values of $\mdot$ but otherwise identical binary parameters. The early time evolution ($t \lesssim 50P$) is similar in all cases, following that described in Section~\ref{sec:bound}.  At later times, however, once the gas falls back to the binary in the $\mdot=10^{-1}\,\myr$ model (as evidenced by a decrease in potential energy), the thermal energy begins to grow, and eventually the total energy approaches a positive value.  Gas falling back in the $\mdot=10^{-3}\,\myr$ model approaches much closer to the central binary than in the higher $\mdot$ runs.  Although the kinetic and thermal energy increase also in this case, the total energy decreases and the gas remains bound.  Most of the kinetic energy comes from motions in the tangential direction (Fig.~\ref{fig:f}, top panel), its total value being about half the absolute value of the potential energy, indicating that the disk is virialized.

The evidence above suggests that the nature of the final configuration is determined by the efficiency with which the bound gas returning to the binary can cool radiatively.  In the limit of efficient cooling, heat deposited in the gas by the binary is immediately radiated, preventing the gas from accumulating enough energy to become unbound.  The result is a circumbinary disk with continuous input of mass and angular momentum at the inner boundary supplied by the \ltwo\ spiral stream, i.e.~an ``excretion disk''.  The evolution of such a disk will eventually be driven by viscous processes.  However, because the only viscosity in our simulations is numerical (the strength of which depends on the resolution), we cannot reliably simulate this subsequent evolution phase. Instead, we refer to previous work on viscous circumbinary disks \citep[e.g.][]{pringle91,bonnell94,rafikov13,rafikov16}. In the opposite limit of inefficient cooling, the gas retains enough thermal energy for pressure gradients to drive an unbound isotropic outflow.  As the heating rate can exceed the Eddington luminosity, convection cannot transport energy outwards efficiently, in which case \citet{quataert16} predict an outflow similar to that we find. The intermediate case, where cooling approximately balances energy deposition, results in the inflated convective envelope (bottom panel of Fig.~\ref{fig:f}).

To test the cooling efficiency hypothesis, we performed otherwise identical simulations but with radiative diffusion and cooling artificially shut off\footnote{ Without radiative cooling and diffusion, the explicit time step becomes much longer and we are able to follow the evolution for a greater number of orbits than in the default calculation.}.  An example is shown with blue lines in Figure~\ref{fig:energy}, where the same combination of binary parameters and $\mdot$ that results in the formation of an excretion disk in the full simulation instead leads to an isotropic outflow when radiative diffusion and cooling are neglected.

To better understand the condition separating outflows from bound disks, we consider the radiative diffusion timescale through the binary envelope,
\beq
\tdiff \sim \frac{\kappa \bar{\rho}R^2}{c},
\label{eq:tdiff}
\eeq
where $\kappa$ is the opacity in the binary envelope of average density $\bar{\rho}$ and size $R \sim 10a$. The advection timescale is defined as that required to replace the mass in the envelope,
\beq
\tadv \sim \frac{4\pi \bar{\rho} R^3}{\mdot}.
\label{eq:tadv}
\eeq
We postulate that the dividing line between unbound outflows and a bound disk is set by the ratio of these two timescales,
\beq
\frac{\tdiff}{\tadv} \sim \frac{\kappa \mdot}{40\pi ac} \approx 0.4 \left(\frac{\kappa}{0.4\,{\rm cm}^2\,{\rm g}^{-1}} \right) \left(\frac{\mdot}{10^{-1}\,\myr}\right) \left(\frac{a}{0.1\,{\rm AU}} \right)^{-1}.
\label{eq:ratio}
\eeq
The assumption of a constant opacity is reasonably for the high-temperature virialized material around the binary (Fig.~\ref{fig:anim_all}), but in detail $\tdiff/\tadv$ will depend also on the binary parameters such as $a$ and $M$ implicitly through the temperature and density dependence of $\kappa$.

\begin{figure}
\centering
\includegraphics[width=\columnwidth]{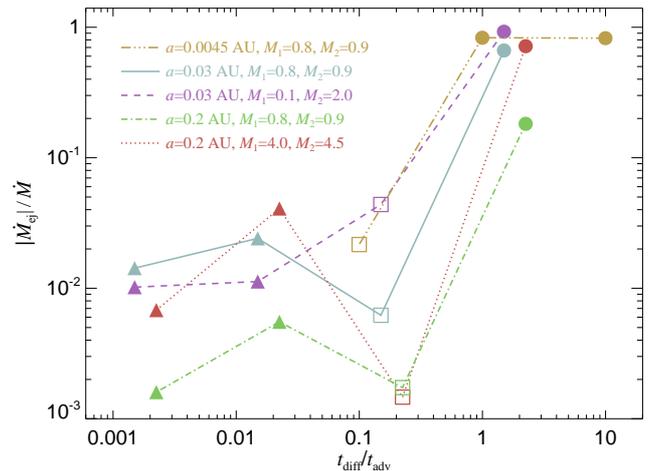}
\caption{\label{fig:t_mdot}Ratio of the time-derivative of the unbound mass, $\mdot_{\rm ej}$, to the mass-loss rate from the binary, $\mdot$, as a function of the ratio of the diffusion and advection timescales (Eq.~[\ref{eq:ratio}]). Different line styles and colors distinguish binary parameters as given in the legend. Filled triangles indicate circumbinary disks, filled circles isotropic winds, and open squares mark simulations ending as an inflated envelope or one of the ambiguous cases shown in Table~\ref{tab:list}. }
\end{figure}

Figure~\ref{fig:t_mdot} shows the mass outflow rate $\mdot_{\rm ej}$ relative to the binary mass-loss rate $\mdot$ as a function of the ratio $\tdiff/\tadv$. $\mdot_{\rm ej}$ is calculated as the time derivative of the total mass of particles for which (1) the sum of the gravitational potential and kinetic energies is positive\footnote{Thermal energy is not included in the estimate of binding energy, because it can be lost to radiation before being transferred to the outflow kinetic energy.  Our estimates of unbound mass thus represent a lower limit.} and (2) the velocity vector points outward.  We evaluate $\mdot_{\rm ej}$ at the end of each simulation, at epochs $t \gtrsim 100P$. Figure~\ref{fig:t_mdot} shows that for $\tdiff/\tadv \gtrsim 1$ the mass of unbound material increases in proportion to the binary mass-loss rate, indicating the presence of an outflow.  By contrast, for the models resulting in an excretion disk or inflated envelope, the asymptotic rate of change of the unbound mass is much smaller, compatible with zero.  By investigating the final flow patterns, we find that the inflated envelope solutions occur for $\tdiff/\tadv \sim 0.1$.  The bottom panel of Figure~\ref{fig:f} shows the flow structure in the best case of an inflated envelope that we can identify.  The scarcity of this outcome in our models implies that the range of $\tdiff/\tadv$ enabling this configuration is narrow, less than an order of magnitude.   Excretion disks form for $\tdiff/\tadv \ll 1$.

\begin{figure*}
\centering
\includegraphics[height=0.4\textwidth]{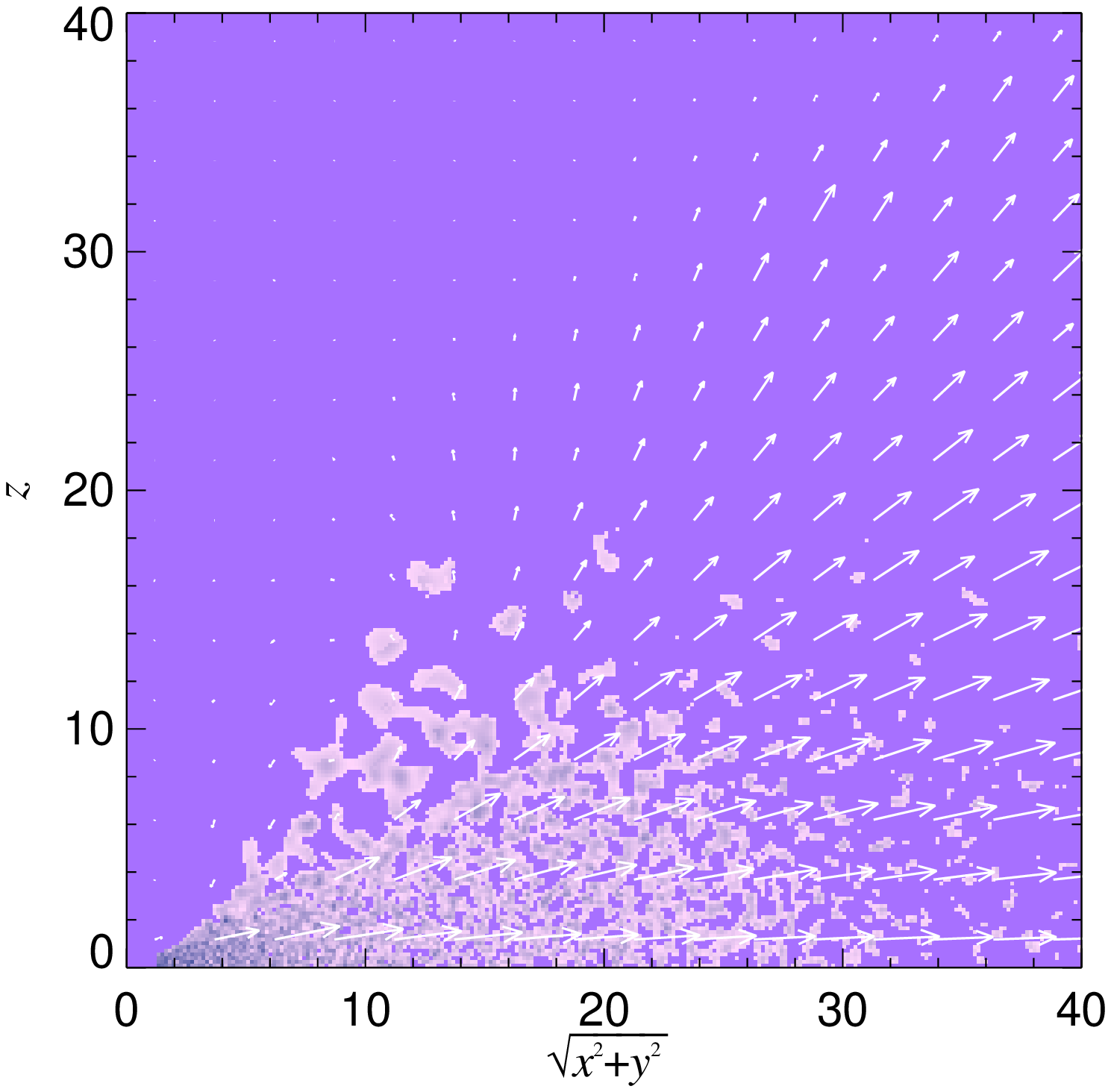}
\includegraphics[height=0.4\textwidth]{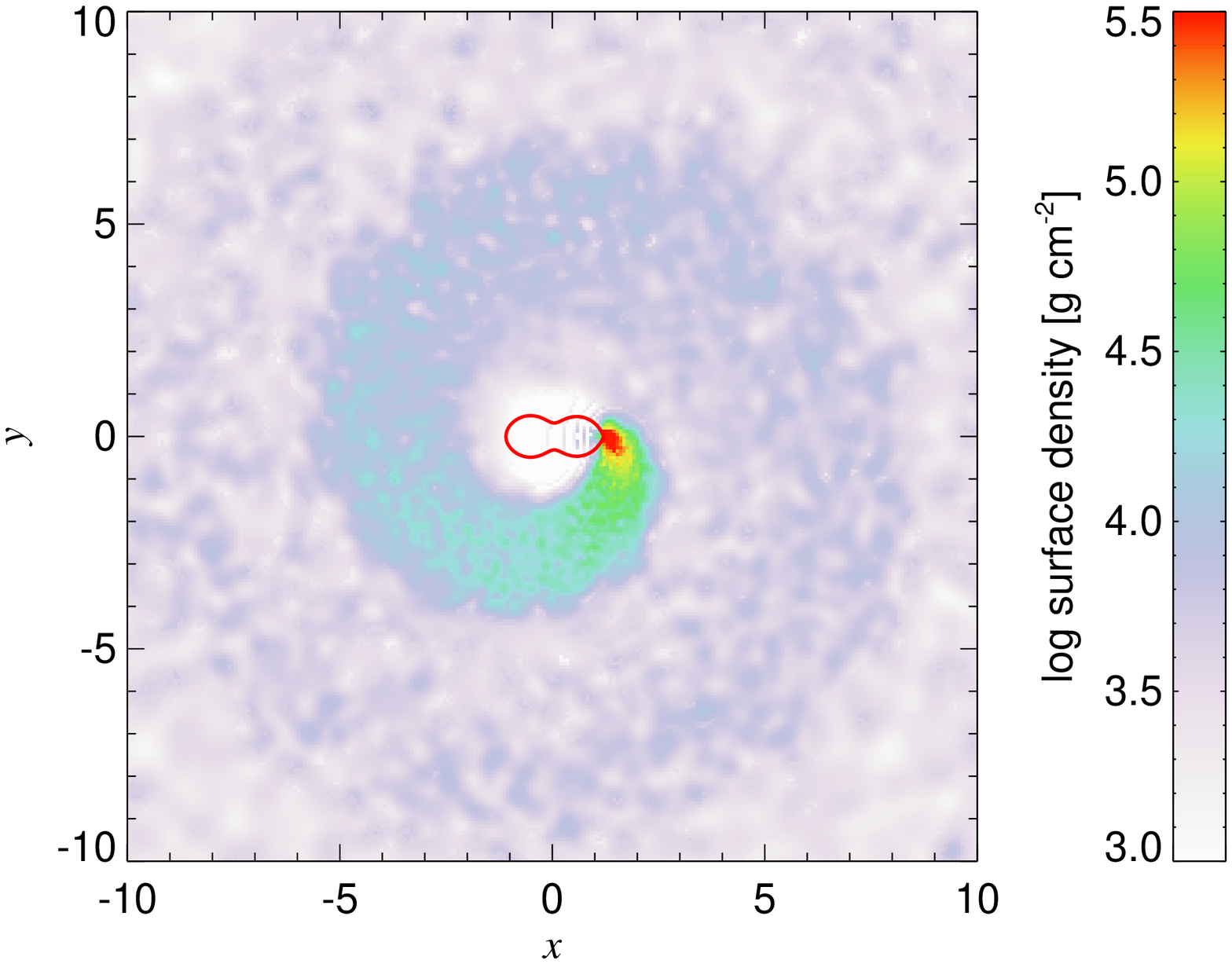}
\caption{\label{fig:f_eq} Final outflow configuration for a binary with $\varepsilon \approx 0.15$ ($a=1.3$\,AU, $M_1=0.8\,\msun$, $M_2=0.9\,\msun$, and $\mdot = 10^{-1}\,\myr$). The left panel shows the vertical temperature structure and velocity field on the same color scale as Figure~\ref{fig:anim_all}, and the right panel shows the surface density of the gas in the vicinity of binary showing a wide stream emanating from \ltwo.}
\end{figure*}

Finally, we have explored simulations for which the value of $\varepsilon$ is higher than that appropriate for unstripped photosphere of the binary.  This case produces a fourth type of final configuration, for which \ltwo\ mass loss almost immediately forms an equatorial outflow with an asymptotic velocity of a quarter to third of $\vesc$, as shown in Figure~\ref{fig:f_eq}. The gas becomes unbound after its initial injection when it is still close to the binary, and there is no fall-back phase as in the cases discussed previously.  This outflow configuration is similar to the equatorial wind which occurs for $0.064 \lesssim q \lesssim 0.78$ analyzed in \citetalias{pmt16}, but we emphasize that since here $q$ lies outside this range, the outflow should be bound according to the criterion of \citet{shu79}. 

To explore this behavior in a different opacity regime, we simulated a binary with $a=0.03$\,AU, $M_1=0.8\,\msun$, $M_2=0.9\,\msun$, and $\mdot=10^{-3}\,\myr$ with $\varepsilon \approx 0.2$, which corresponds to an injection temperature at \ltwo\ of about $10^5$\,K. We obtained a similar outflow to that shown in Figure~\ref{fig:f_eq}, even though an otherwise identical model with a lower value of $\varepsilon = 0.05$ instead produces an excretion disk.  In general, we find that for $\varepsilon \gtrsim 0.15$ the differential tidal torqueing by the binary is more efficient.  This causes the spiral streams to collide and heat much closer to the binary, allowing the gas to become unbound even when the binary mass ratio lies outside of the nominal range for an unbound outflow according to \citet{shu79} and \citetalias{pmt16}.  We emphasize that the gas ejected at \ltwo\ is initially bound, even for our equation of state, which takes into account ionization and molecular energies.  We find similar results when repeating the same calculation using a simple polytropic EOS, for which internal energy is significantly smaller in the temperature and density range of interest. 

Equatorial winds with $\varepsilon\gtrsim 0.15$ are significantly more radiatively efficient than the similar outflows we analyzed in \citetalias{pmt16}. For example, a binary with $M_1=0.225\,\msun$, $M_2=1.5\,\msun$, $a=0.03\,AU$, $\mdot = 10^{-3}\,\myr$, and $\varepsilon =0.05$ reaches an asymptotic luminosity of only $\sim 30\,\lsun$ \citepalias{pmt16}.  By contrast, here we find for identical $a$, $\mdot$, similar total mass ($M_1=0.8\,\msun$, $M_2=0.9\,\msun$), and $\varepsilon \approx 0.2$ that the luminosity reaches a much higher value of $\sim 2000\,\lsun$. In \citetalias{pmt16}, we found that $L$ is increasing with $\varepsilon$ when other parameters are held fixed (Fig.~B1), but here we extend the calculations to higher values of $\varepsilon$.

\subsection{Radiative properties}
\label{sec:radiative}

\begin{figure}
\centering
\includegraphics[width=\columnwidth]{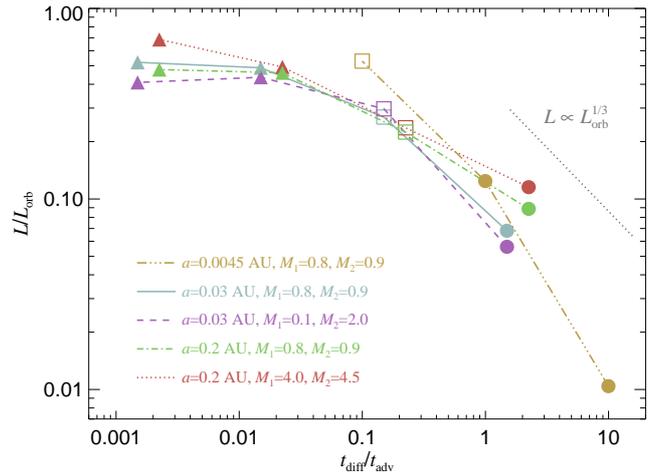}
\caption{\label{fig:t_lum} Luminosity of the binary outflow as a function of the ratio of diffusion to advection timescales. Meaning of the symbols is the same as in Fig.~\ref{fig:t_mdot}. The gray dotted line shows the expected scaling of the super-Eddington wind luminosity \citep{quataert16}. }
\end{figure}

A natural scale for the luminosity of the merger ejecta is given by
\beq
\lorb = \frac{\mdot \vorb^2}{2} \approx 1.5\times 10^4\,\lsun\ \left(\frac{\mdot}{10^{-2}\,\myr}\right) \left(\frac{M}{2\,\msun}\right)\left(\frac{a}{0.1\,{\rm AU}}\right)^{-1}.
\label{eq:lorb}
\eeq
Figure~\ref{fig:t_lum} shows the radiative efficiency $L/\lorb$ measured in each simulation as a function of $\tdiff/\tadv$.  The excretion disk solutions radiate about half of $\lorb$, while the radiation efficiencies decrease to $\sim 20\%$ for the inflated envelopes and drop to $\lesssim 10\%$ for the outflows.  A high value of $\tdiff/\tadv$ implies that the gas must expand to radiate, resulting in a loss of energy to adiabatic expansion. \citet{quataert16} predicted that the luminosity emerging from super-Eddington winds should scale with the heating rate as $L \propto \lorb^{1/3}$. For constant $a$ and $\kappa$, this implies $L/\lorb \propto \mdot^{-2/3}$, which we show in Figure~\ref{fig:t_lum}. Our results are not inconsistent with this prediction, but calculations of the super-Eddington wind with more realistic opacities are needed for a more detailed comparison.

\begin{figure}
 \centering
\includegraphics[width=\columnwidth]{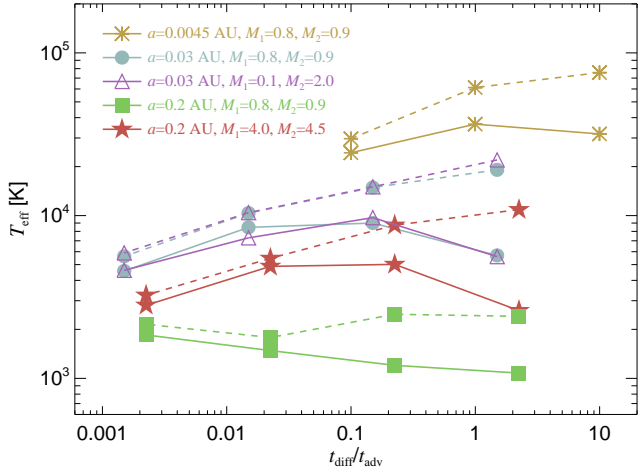}
\caption{\label{fig:t_teff} Estimates of the temperature of the radiation $\teff$ as a function of the ratio of diffusion to advection timescales. Solid lines show the final temperatures after $\gtrsim 100P$ of evolution, while the dashed lines show maximum $\teff$ for each binary.}
\end{figure}

Figure~\ref{fig:t_teff} shows the effective temperature of the emission, $\teff$, for each solution.  The temperature depends only weakly on $\tdiff/\tadv$, except for a noticable decrease in $\teff$ for $\tdiff/\tadv \gtrsim 1$, which corresponds to the establishment of recombination front in the outflow (see also Fig.~\ref{fig:lc}). The value of $\teff$ is more sensitive to the orbital speed of the binary, showing an increase with $v_{\rm orb}$ similar to that found in \citetalias{pmt16}.  We caution, however, that our estimates of $\teff$ are less reliable than our estimates of the luminosity.

A more detailed description of the radiative properties of the isotropic outflow (and to a lesser extent of the equatorial outflow) could be provided by a one-dimensional steady-state wind calculation with a radiative transport such as flux-limited diffusion.  Indeed, the resulting luminosities and effective temperatures of winds with super-Eddington energy deposition near the inner boundary have broader applicability to a range of astrophysical environments, such as luminous blue variables and classical novae  \citep{quataert16,shen16}. We defer such a calculation to future work.

Finally, we note that the outflows studied in \citetalias{pmt16} were radiatively inefficient compared to those described here. The highest luminosities achieved for optically-thin outflows were found to be $\sim \mdot (\Delta v)^2/2$, where $\Delta v \approx 0.08\vorb$ is the spread in the outflow velocity induced by binary torques, leading to luminosities of at most $\sim 7\times 10^{-3}\,\lorb$. 

\subsection{Dependence on binary parameters}

Marginally-bound outflows occur for $q \gtrsim 0.78$ or $q \lesssim 0.064$ \citep{shu79}. To reveal any potential differences between these two mass ratio regimes, Figures~\ref{fig:t_mdot}, \ref{fig:t_lum}, and \ref{fig:t_teff} show results for a binary with $q=0.05$ and a total mass of $M = 2.1\,\msun$.  The results are very similar to a binary with $M=1.7\,\msun$ and $q =0.89$, indicating that these two regimes are essentially indistinguishable. Nonetheless, we expect that the tidal torquing will become inefficient for $q \rightarrow 0$.  Such extreme mass ratios are not relevant to realistic stellar binaries, but could describe planet-star mergers \citep{metzger12}.

\subsection{Backreaction on the binary orbit}

\begin{figure}
\centering
\includegraphics[width=\columnwidth]{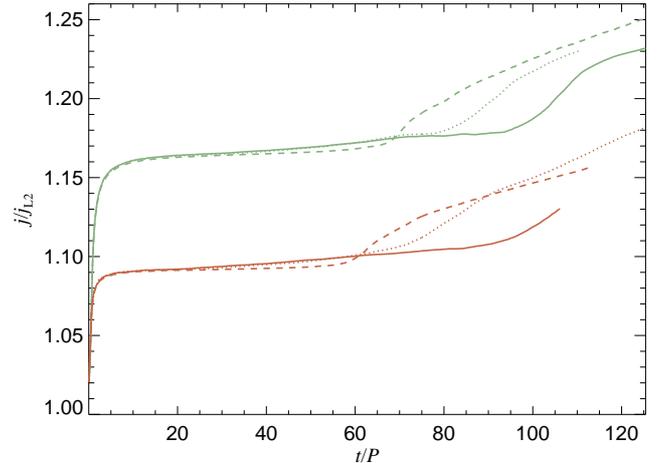}
\caption{\label{fig:ang_mom} Specific angular momentum of the active particles in the simulation relative to the specific angular momentum of the \ltwo\ point as a function of time. We show results for two binaries with identical semi-major axis $a=0.03$\,AU, but different mass ratios: $M_1=0.8\,\msun$ and $M_2=0.9\,\msun$ (green lines), and $M_1=0.1\,\msun$ and $M_2=2.0\,\msun$ (red lines). For each of the binaries we show mass ratios $\mdot=10^{-1}\,\myr$ (solid lines), $10^{-2}$ (dotted), and $10^{-3}$ (dashed).}
\end{figure}

Although we do not evolve the properties of the central binary, we can estimate the backreaction on the binary orbit by following the evolution of angular momentum and energy of the gas in the simulation. In Figure~\ref{fig:ang_mom}, we show the time evolution of the specific angular momentum calculated as a ratio of total angular momentum of all active particles $|\mathbf{J}|$ to the total active mass in the simulatins. We relate this quantity to the specific angular momentum of the \ltwo\ point, $j_{{\rm L}_2}$. We find that shortly after the simulation commences, the specific angular momentum increases due to tidal torqueing from the binary by several tens of percent, depending on the binary mass ratio \citepalias{pmt16}. When enough gas falls back to the binary, the specific angular momentum starts increasing again as tidal torques affect more material. There is no qualitative difference between the three final configurations with $\varepsilon \ll 1$. For the fourth case of equatorial wind, the final rise of angular momentum is not present.

The ultimate reservoir supplying energy to the gas is the orbit of the central binary star, and we would like to estimate this energy drain by the processes described here and in \citetalias{pmt16}.  Since we do not include the structure of the individual stars, we can only discuss the behavior of test masses.  The asymptotic energy of any mass loss is eventually dominated by its kinetic energy, and we showed in Figure~\ref{fig:vel_theta} and in \citetalias{pmt16} that the asymptotic velocity for these outflows $v_{\infty}$ is invariably about quarter to third of $\vesc$.  As $v_{\infty}^{2}$ is small compared to the potential energy at \ltwo\ or the \ltwo\ corotation kinetic energy, we conclude that most of the binary energy is expended in unbinding the gas instead of contributing to its asymptotic kinetic energy.  Note that the \ltwo\ corotation energy is comparable, but always smaller than, the potential energy.  Bringing the gas to corotation at \ltwo\ itself requires energy and this probably comes also from the orbit of the binary.

\subsection{Dependence on resolution and inner boundary condition}

\begin{figure*}
\includegraphics[width=0.49\textwidth]{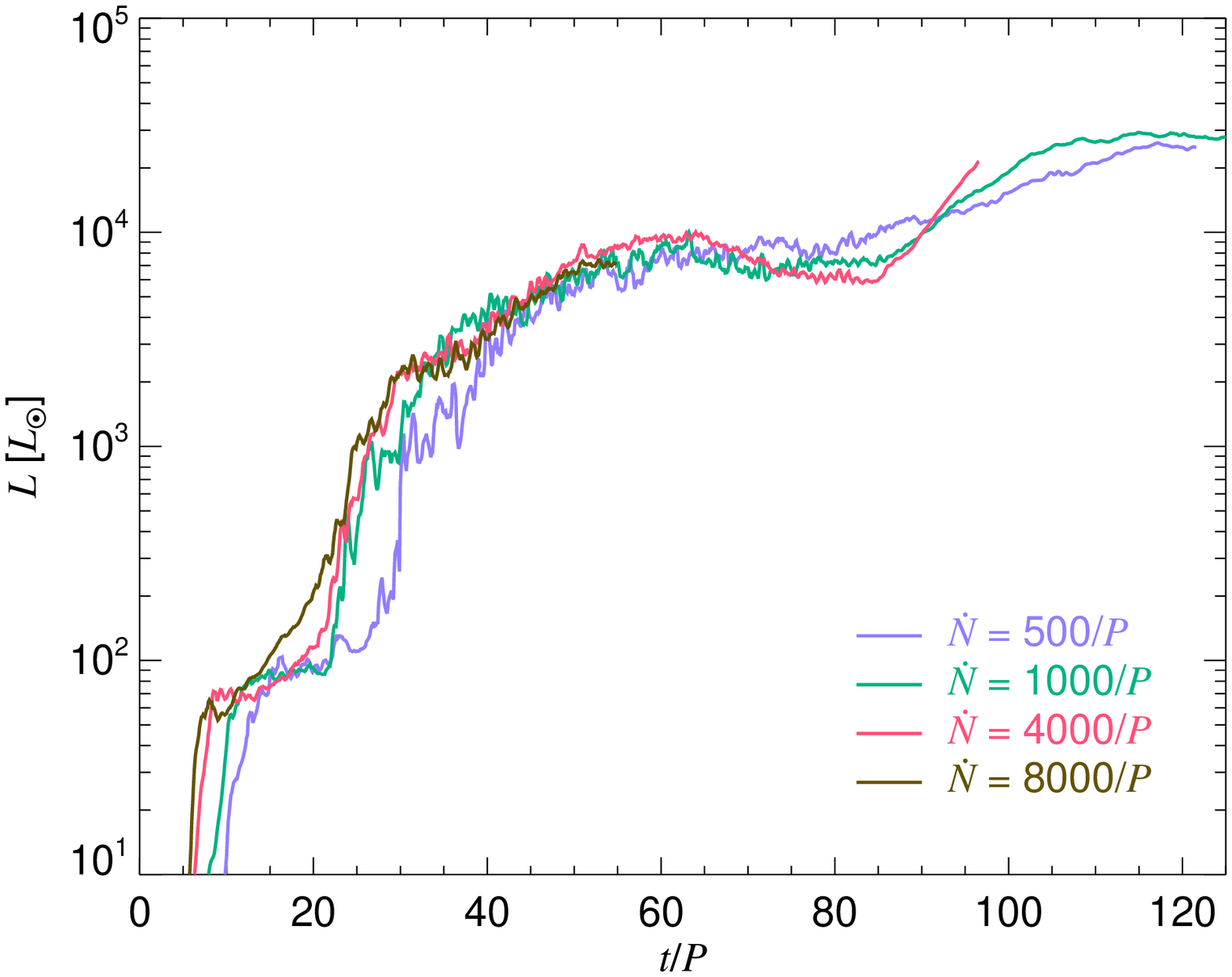}\includegraphics[width=0.49\textwidth]{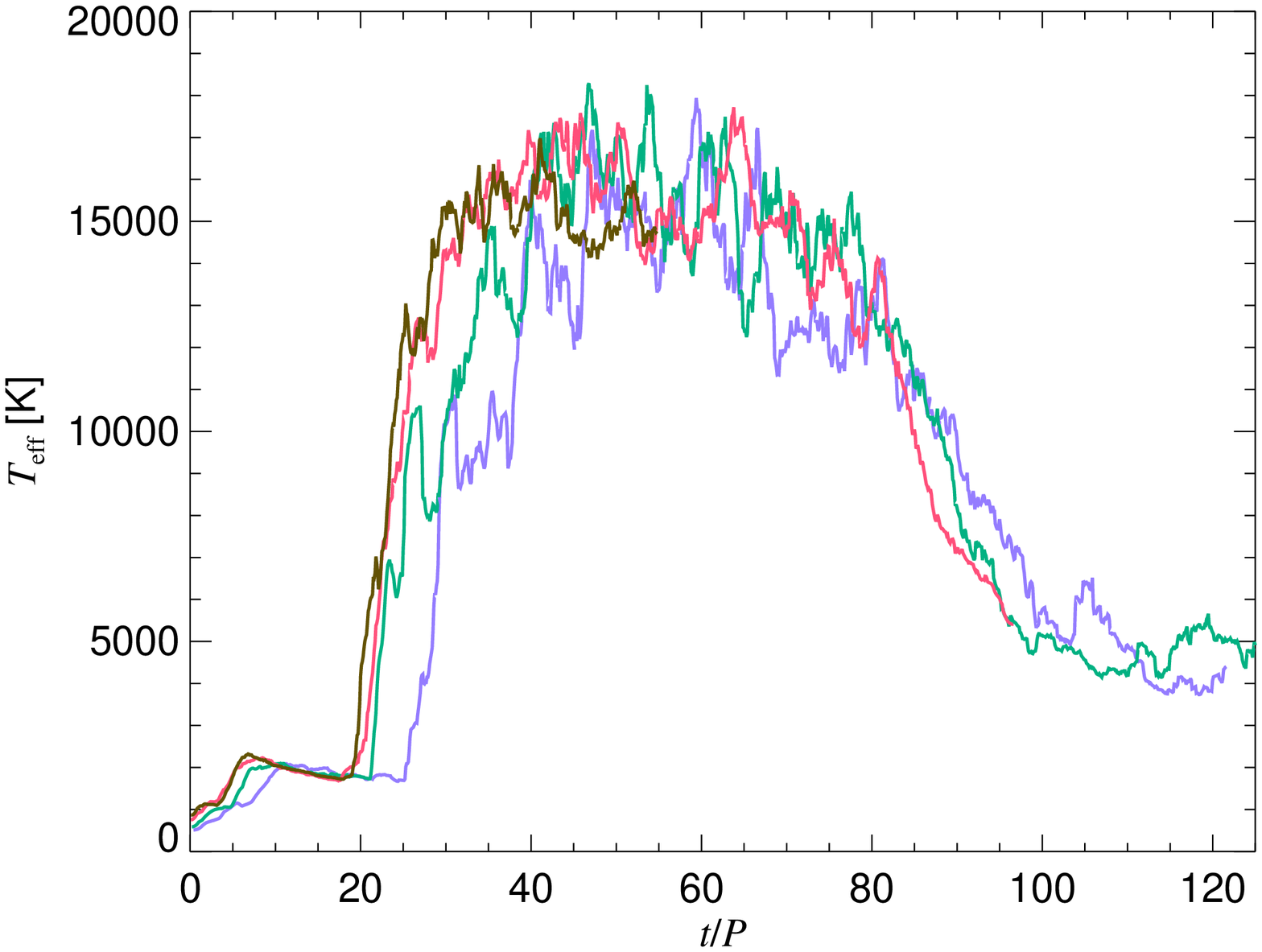}
\caption{\label{fig:res} Dependence of our results on the number of particles for a binary with $a=0.03$\,AU, $M_1=0.8\,\msun$, $M_2=0.9\,\msun$, and $\mdot = 10^{-1}\,\myr$. We show four different values of $\dot{N}$ both higher and lower than our default $\dot{N} = 1000/P$. }
\end{figure*}

A proper resolution study is difficult, because of the necessity of simulating for $\sim 100P$ to reach the final configuration. In Figure~\ref{fig:res}, we show a comparison of luminosity and effective temperature evolution for simulations with different rate of particle injection $\dot{N}$. Both lower and higher resolution than the default yield similar results, albeit higher resolutions typically produce sharper features in the luminosity evolution and lower scatter.  This implies that the adopted resolution is sufficient to qualitatively capture the results. 

Figure~\ref{fig:lc} shows with dashed lines calculations which instead employ a reflective inner boundary. These runs typically produce slightly higher luminosities and effective temperatures, presumably because the hot virialized gas is not absorbed by the inner boundary. Nonetheless, the evolution is not qualitatively different from our runs with an absorbing inner boundary, and the quantitative results are similar as well. The exact treatment of the inner boundary is not overly important, because the gas returns with considerable angular momentum and interacts with the binary itself relatively weakly.

\section{Summary of outcomes of \ltwo\ mass loss}
\label{sec:summary}

\begin{figure}
\includegraphics[width=\columnwidth]{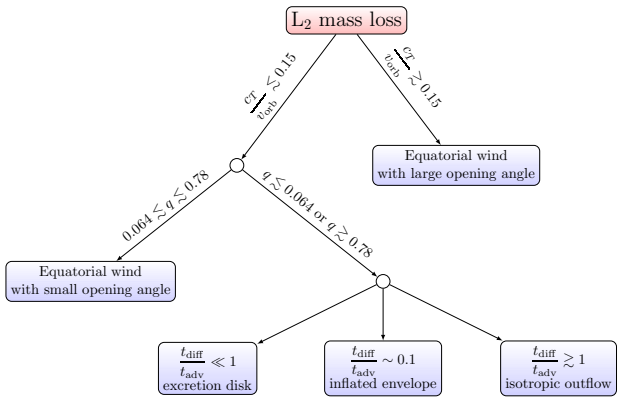}
\caption{\label{fig:dec} Decision tree for determining the final configuration of the \ltwo\ mass loss as a function of $\varepsilon \equiv c_T/\vorb$, binary mass ratio $q$, and the radiative cooling efficiency $\tdiff/\tadv$.}
\end{figure}

Here we summarize the results from \citetalias{pmt16} and this paper on radiation-hydrodynamics of outflows from \ltwo. The range of possible outcomes and the phenomenology is richer than what was envisioned in the analytic calculation of \citet{shu79}. The subsequent presentation assumes that the material is nearly corotating at \ltwo\ and the evolution proceeds gradually on a timescale of at least tens of orbits. The results are summarized graphically in Figure~\ref{fig:dec}.

If the thermal content of the gas at \ltwo\ is sufficiently high, the result is an equatorial wind irrespective of the binary mass ratio, because the spread of the stream allows for more efficient transfer of energy from the central binary. Specifically, this happens if the ratio of the sound speed to the orbital velocity is $\gtrsim 0.15$. This is also approximately the vertical opening angle of the outflow (Fig.~\ref{fig:f_eq}). The resulting outflow exhibits internal shocks and might form dust, depending on the parameters of the binary. This is probably the final stage in the evolution of any merger, because as more surface layers of the mass-lossing star are lost, $\varepsilon$ steadily increases. The radiated luminosity depends on the optical depth through the outflow, but can achieve tens of percent of $\lorb$, which is much higher than the outflows investigated in \citetalias{pmt16}.

If the thermal content of the gas is low ($\varepsilon \lesssim 0.15$), then the evolution depends on the binary mass ratio. For mass ratios $0.064 \lesssim q \lesssim 0.78$, the spiral stream merges and forms an equatorial outflow \citepalias{pmt16}. The merging process thermalizes $\sim 5\%$ of the kinetic energy of the outflow, which is roughly constant as a function of binary parameters. As a result, the outflow is radiatively very inefficient when measured relative to $\lorb$. Dust can form in copious quantities. 

\begin{figure}
 \centering
\includegraphics[width=0.45\textwidth]{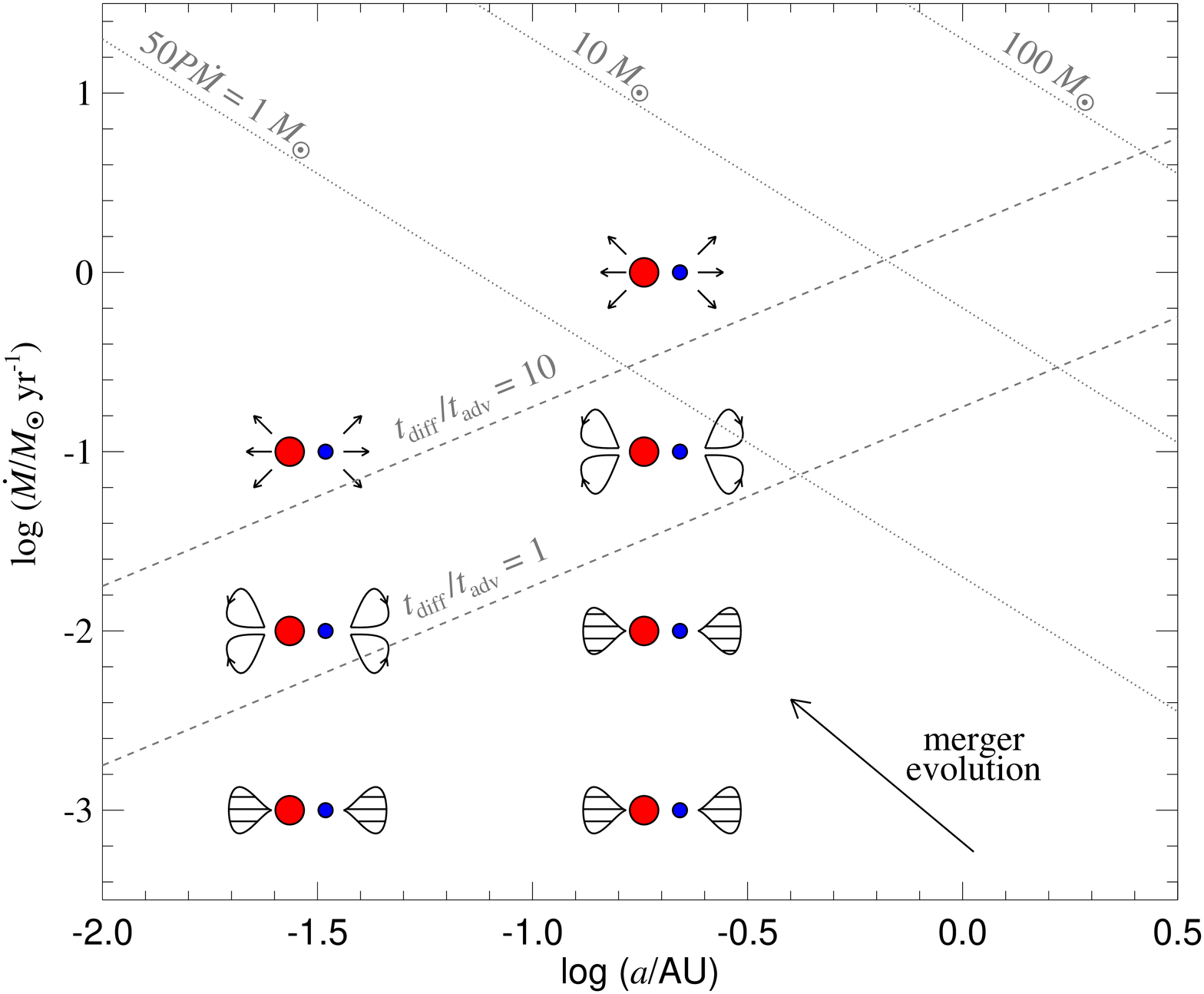}
\caption{\label{fig:table} Summary of gas configurations for \ltwo\ mass loss with $\varepsilon \lesssim 0.15$, binary mass ratios $q \gtrsim 0.78$ or $q \lesssim 0.064$, and the assumption that the gas is in corotation at \ltwo. The three types of symbols from bottom to top denote excretion disks, inflated envelopes, and isotropic outflows and their positions are based directly on the results of our simulations. The dashed diagonal lines show approximate dividing lines between these configurations based on Eq.~(\ref{eq:ratio}). Solid diagonal lines mark the maximum $\mdot$ that can be sustained by a binary with given total mass $M$ and semi-major axis $a$ for $50$ orbital periods.}
\end{figure}

The dynamics of cool ($\varepsilon \lesssim 0.15$) \ltwo\ mass loss for binaries with $q \gtrsim 0.78$ or $q \lesssim 0.064$ was the main subject of this paper. We find that for a few tens of orbits after the initiation of the \ltwo\ mass loss the binary produces a relatively low-temperature low-luminosity outflow described in \citetalias{pmt16}. Then, a progressively larger fraction of the narrow equatorial outflow stalls and falls back to the central binary. Near the binary, the gas is shocked and its motion is randomized resulting in a hot envelope virialized with the binary orbit. As a result, the object will slowly brighten on a fallback timescale, which can be about a hundred binary orbital periods. This brightening will be driven by an increase in the effective temperature at constant or shrinking radius, if all physical parameters are held fixed (Fig.~\ref{fig:lc}). A small fraction of the heated gas escapes along the poles, where it is not blocked by the dense equatorial ring of fallback material. As the equatorial ring falls back and joins the virialized envelope, three outcomes are possible depending on the efficiency of radiative cooling of the virialized envelope parameterized by the ratio of diffusion to advection timescale. If the cooling is inefficient, the barrier in the orbital plane eventually disappears and the binary develops a nearly isotropic wind driven by super-Eddington energy deposition \citep{quataert16} around the binary (Fig.~\ref{fig:anim_all}). If the base of the wind is sufficiently hot, the photosphere in the wind will be at the hydrogen recombination front with  $\teff \sim 5000$\,K. If the cooling is very efficient, the heating from the binary is radiated away and the outflow collapses to an excretion disk, which is fed with mass and angular momentum at the inner edge (Fig.~\ref{fig:f}, top panel). The intermediate case results in a cool inflated envelope, where a global meridional circulation brings the binary heating to the surface, where it can be efficiently radiated (Fig.~\ref{fig:f}, bottom panel). In all of these cases the outflow can radiate $10^{-2}$ to $\sim$few $10^{-1}$ of $\lorb$.

During a merger event, the mass loss rate increases and the binary separation decreases, which drives an increase in the ratio of diffusion to advection timescales (Eq.~[\ref{eq:ratio}]). If $\varepsilon$ stays $\lesssim 0.15$ throughout this evolution and the binary mass ratio does not change considerably, the merger should evolve from displaying an excretion disk to an isotropic wind, as schematically indicated in Figure~\ref{fig:table}. If excretion disk is the ultimate final state, a small amount of the mass will bring most of the angular momentum will outward on the viscous time, while most of the mass will remain near the binary.

\section{Implications for red transients}
\label{sec:implications}

Our results provide a natural explanation for prolonged pre-maximum activity of stellar mergers through the action of \ltwo\ mass loss, which slowly prepares the binary for the ultimate dynamical event. Gradual brightening before the main peak can result either from a slow increase in $\mdot$, which progressively shifts the photospheric radius outward at approximately constant effective temperature, as was suggested for V1309~Sco \citep{pejcha14}, or by the launching of marginally-bound ejecta, which later returns and interacts with the binary.  In the latter case, the photospheric radius should initially remain approximately constant, while the effective temperature increases. The transition from a bound hot virialized envelope to a colder isotropic wind brings with it a sharp increase, by up to an order of magnitude, in the luminosity. With some tuning of the initial conditions, it is also possible to get an inflated convective envelope, which should lie near the Hayashi track, as was argued to describe one phase in the evolution of V838~Mon \citep{evans03,tylenda05}. Despite the richness of diversity it allows, \ltwo\ mass loss may be only a part of the story\footnote{Accretion energy and the associated possibility of jets might play a role in some red transients \citep{akashi15,kashi15}. \citet{macleod16} argued dynamically-driven ejecta at the onset of a common envelope episode could explain the red transient M31 LRN 2015.}.  Without including the time-changing properties of the binary and its mass loss rate, it is hard to draw a complete and consistent picture of individual transients. We defer such an application to future work.

\citetalias{pmt16}  predicted a correlation in RT between the peak luminosity and the expansion velocity estimated from the widths of Balmer lines.  However, the luminosities of the outflows studied in \citetalias{pmt16} were too low at a given expansion velocity to explain the full range of observed transients. By contrast, the winds investigated here consistently show much higher radiative efficiencies, resulting in higher luminosities for a given terminal velocity (Sec.~\ref{sec:radiative}).  These higher luminosities are now sufficient to match the observed correlation from \citetalias{pmt16}, suggesting that the majority of RT may be better explained by winds with high $\varepsilon$ or from gas virialized with the binary orbit.

Both this work and \citetalias{pmt16} predict that \ltwo\ mass loss can emerge as a wind. The resulting ejecta velocity distribution is markedly different than that from an explosion, which instead produces a homologuous structure with slower matter positioned inside of faster material.  In an explosion, as the pseudo-photosphere recedes in mass coordinate, the expansion velocity as measured from the P Cyg absorption profile decreases with time, as is observed in Type II-Plateau supernovae. By contrast, in a wind the pseudo-photosphere is positioned outside of the sonic point and hence a constant expansion velocity is measured with time, or perhaps the line widths may even {\em increase} as the binary separation shrinks and the wind velocity rises towards merger.  The homologous assumption being invalid for winds calls into question whether scaling relations for luminosities and transient durations can be transferred from the context of core-collapse supernovae to RT \citep{ivanova13b}. Distinguishing between a slowly-evolving wind and an explosion ejecta geometry may be possible through careful spectral monitoring of the outburst.

We conclude by speculating about a possible backreaction of the circumbinary gas on the binary mass transfer rate. The final frame of Figure~\ref{fig:anim_all} shows that the central binary is surrounded by gas virialized with respect to the binary orbit. Due to centrifugal barrier (and depending on the exact behavior of the gas near the inner boundary), the binary itself might reside in a cavity devoid of gas. This cavity will be soon filled with radiation of a temperature $T \sim 10^5$\,K, which is much higher than that of the binary surface, even if some of the surface layers have been stripped.  Initially, the resulting radiation pressure should dynamically slightly compress the surface of the binary. Over the diffusion time of the surface layers, the star may absorb some of this energy, causing the expansion and evaporation of stellar material. Since the physics of the surface layers plays a crucial role in (at least the very early) evolution of the merger, this effect might affect the rate of mass transfer or mass loss. The evolution of stars in a radiative bath has been considered previously in the context of active galactic nuclei and X-ray binaries \citep[e.g.][]{tout89,podsiadlowski91,harpaz91}, but to our knowledge has not yet been considered in the context of merging stars.  Future work is necessary to explore the importance of this effect. 

\section*{acknowledgements}
We thank the anonymous referee for constructive comments that improved the paper. We thank Chris Kochanek for detailed reading of the manuscript. OP acknowledges discussions with Roman Rafikov, Jeremy Goodman, and Eliot Quataert. The simulations were carried out using computers supported by the Princeton Institute of Computational Science and Engineering. Support for OP was provided by NASA through Hubble Fellowship grant HST-HF-51327.01-A awarded by the Space Telescope Science Institute, which is operated by the Association of Universities for Research in Astronomy, Inc., for NASA, under contract NAS 5-26555. BDM acknowledges support from NSF grant AST-1410950, NASA grants NNX15AR47G and NNX16AB30G, and the Alfred P. Sloan Foundation.


\begin{thebibliography}{}
\footnotesize
\bibitem[Abbott et al.(2016a)]{abbott16a} Abbott, B.~P., Abbott, R., Abbott, T.~D., et al.\ 2016a, Physical Review Letters, 116, 061102 
\bibitem[Abbott et al.(2016b)]{abbott16b} Abbott, B.~P., Abbott, R., Abbott, T.~D., et al.\ 2016b, \apjl, 818, L22 
\bibitem[Akashi et al.(2015)]{akashi15} Akashi, M., Sabach, E., Yogev, O., \& Soker, N.\ 2015, \mnras, 453, 2115
\bibitem[Balsara(1995)]{balsara95} Balsara, D.~S.\ 1995, Journal of Computational Physics, 121, 357 
\bibitem[Belczynski et al.(2016)]{belczynski16} Belczynski, K., Holz, D.~E., Bulik, T., \& O'Shaughnessy, R.\ 2016, arXiv:1602.04531 
\bibitem[Bodenheimer et al.(1990)]{bodenheimer90} Bodenheimer, P., Yorke, H.~W., Rozyczka, M., \& Tohline, J.~E.\ 1990, \apj, 355, 651 
\bibitem[Bond et al.(2003)]{bond03} Bond, H.~E., Henden, A., Levay, Z.~G., et al.\ 2003, \nat, 422, 405 
\bibitem[Bonnell \& Bate(1994)]{bonnell94} Bonnell, I.~A., \& Bate, M.~R.\ 1994, \mnras, 269,  
\bibitem[Colagrossi \& Landrini(2003)]{colagrossi03} Colagrossi, A., \& Landrini, M.\ 2003, Journal of Computational Physics, 191, 448 
\bibitem[Evans et al.(2003)]{evans03} Evans, A., Geballe, T.~R., Rushton, M.~T., et al.\ 2003, \mnras, 343, 1054 
\bibitem[Ferguson et al.(2005)]{ferguson05} Ferguson, J.~W., Alexander, D.~R., Allard, F., et al.\ 2005, \apj, 623, 585
\bibitem[Forgan et al.(2009)]{forgan09} Forgan, D., Rice, K., Stamatellos, D., \& Whitworth, A.\ 2009, \mnras, 394, 882 
\bibitem[Ge et al.(2010)]{ge10} Ge, H., Hjellming, M.~S., Webbink, R.~F., Chen, X., \& Han, Z.\ 2010, \apj, 717, 724 
\bibitem[Ge et al.(2015)]{ge15} Ge, H., Webbink, R.~F., Chen, X., \& Han, Z.\ 2015, \apj, 812, 40 
\bibitem[Harpaz \& Rappaport(1991)]{harpaz91} Harpaz, A., \& Rappaport, S.\ 1991, \apj, 383, 739 
\bibitem[Herant(1994)]{herant94} Herant, M.\ 1994, \memsai, 65, 1013 
\bibitem[Hjellming \& Webbink(1987)]{hjellming87} Hjellming, M.~S., \& Webbink, R.~F.\ 1987, \apj, 318, 794 
\bibitem[Hwang et al.(2015)]{hwang15} Hwang, J., Lombardi, J.~C., Jr., Rasio, F.~A., \& Kalogera, V.\ 2015, \apj, 806, 135 
\bibitem[Ivanova et al.(2013)]{ivanova13b} Ivanova, N., Justham, S., Avendano Nandez, J.~L., \& Lombardi, J.~C.\ 2013, Science, 339, 433 
\bibitem[Kashi \& Soker(2016)]{kashi15} Kashi, A., \& Soker, N.\ 2016, Research in Astronomy and Astrophysics, 16, 014 
\bibitem[Katz \& Dong(2012)]{katz12} Katz, B., \& Dong, S.\ 2012, arXiv:1211.4584
\bibitem[Kochanek et al.(2014)]{kochanek14} Kochanek, C.~S., Adams, S.~M., \& Belczynski, K.\ 2014, \mnras, 443, 1319 
\bibitem[Kuiper(1941)]{kuiper41} Kuiper, G.~P.\ 1941, \apj, 93, 133 
\bibitem[Kulkarni et al.(2007)]{kulkarni07} Kulkarni, S.~R., Ofek, E.~O., Rau, A., et al.\ 2007, \nat, 447, 458 
\bibitem[Kurtenkov et al.(2015)]{kurtenkov15} Kurtenkov, A.~A., Pessev, P., Tomov, T., et al.\ 2015, \aap, 578, L10 
\bibitem[Lai et al.(1993)]{lai93} Lai, D., Rasio, F.~A., \& Shapiro, S.~L.\ 1993, \apjl, 406, L63 
\bibitem[Lai et al.(1994a)]{lai94a} Lai, D., Rasio, F.~A., \& Shapiro, S.~L.\ 1994a, \apj, 420, 811
\bibitem[Lai et al.(1994b)]{lai94b} Lai, D., Rasio, F.~A., \& Shapiro, S.~L.\ 1994b, \apj, 423, 344 
\bibitem[Lai et al.(1994c)]{lai94c} Lai, D., Rasio, F.~A., \& Shapiro, S.~L.\ 1994c, \apj, 437, 742 
\bibitem[Li \& Zhang(2006)]{li06} Li, L., \& Zhang, F.\ 2006, \mnras, 369, 2001 
\bibitem[Libersky et al.(1993)]{libersky93} Libersky, L.~D., Petschek, A.~G., Carney, T.~C., Hipp, J.~R., \& Allahdadi, F.~A.\ 1993, Journal of Computational Physics, 109, 67 
\bibitem[Lombardi et al.(2011)]{lombardi11} Lombardi, J.~C., Jr., Holtzman, W., Dooley, K.~L., et al.\ 2011, \apj, 737, 49 
\bibitem[Lombardi et al.(2015)]{lombardi15} Lombardi, J.~C., McInally, W.~G., \& Faber, J.~A.\ 2015, \mnras, 447, 25 
\bibitem[MacLeod et al.(2016)]{macleod16} MacLeod, M., Macias, P., Ramirez-Ruiz, E., et al.\ 2016, arXiv:1605.01493 
\bibitem[Martini et al.(1999)]{martini99} Martini, P., Wagner, R.~M., Tomaney, A., et al.\ 1999, \aj, 118, 1034 
\bibitem[Mason et al.(2010)]{mason10} Mason, E., Diaz, M., Williams, R.~E., Preston, G., \& Bensby, T.\ 2010, \aap, 516, A108 
\bibitem[Metzger et al.(2012)]{metzger12} Metzger, B.~D., Giannios, D., \& Spiegel, D.~S.\ 2012, \mnras, 425, 2778 
\bibitem[Monaghan \& Gingold(1983)]{monaghan83} Monaghan, J.~J., \& Gingold, R.~A.\ 1983, Journal of Computational Physics, 52, 374 
\bibitem[Munari et al.(2002)]{munari02} Munari, U., Henden, A., Kiyota, S., et al.\ 2002, \aap, 389, L51 
\bibitem[Nandez et al.(2014)]{nandez14} Nandez, J.~L.~A., Ivanova, N., \& Lombardi, J.~C., Jr.\ 2014, \apj, 786, 39 
\bibitem[Passy et al.(2012)]{passy12} Passy, J.-C., Herwig, F., \& Paxton, B.\ 2012, \apj, 760, 90 
\bibitem[Pavlovskii \& Ivanova(2015)]{pavlovskii15} Pavlovskii, K., \& Ivanova, N.\ 2015, \mnras, 449, 4415 
\bibitem[Pejcha et al.(2013)]{pejcha13} Pejcha, O., Antognini, J.~M., Shappee, B.~J., \& Thompson, T.~A.\ 2013, \mnras, 435, 943
\bibitem[Pejcha et al.(2016)]{pmt16} Pejcha, O., Metzger, B.~D., \& Tomida, K.\ 2016, \mnras, 455, 4351 
\bibitem[Pejcha(2014)]{pejcha14} Pejcha, O.\ 2014, \apj, 788, 22 
\bibitem[Podsiadlowski(1991)]{podsiadlowski91} Podsiadlowski, P.\ 1991, \nat, 350, 136 
\bibitem[Price \& Monaghan(2007)]{price07} Price, D.~J., \& Monaghan, J.~J.\ 2007, \mnras, 374, 1347
\bibitem[Pringle(1991)]{pringle91} Pringle, J.~E.\ 1991, \mnras, 248, 754 
\bibitem[Quataert et al.(2016)]{quataert16} Quataert, E., Fern{\'a}ndez, R., Kasen, D., Klion, H., \& Paxton, B.\ 2016, \mnras, 458, 1214
\bibitem[Rafikov(2013)]{rafikov13} Rafikov, R.~R.\ 2013, \apj, 774, 144 
\bibitem[Rafikov(2016)]{rafikov16}  Rafikov, R.~R.\ 2016, arXiv:1604.07439
\bibitem[Rasio(1995)]{rasio95_wuma} Rasio, F.~A.\ 1995, \apjl, 444, L41 
\bibitem[Rasio \& Shapiro(1992)]{rasio92} Rasio, F.~A., \& Shapiro, S.~L.\ 1992, \apj, 401, 226 
\bibitem[Rasio \& Shapiro(1994)]{rasio94} Rasio, F.~A., \& Shapiro, S.~L.\ 1994, \apj, 432, 242 
\bibitem[Rasio \& Shapiro(1995)]{rasio95} Rasio, F.~A., \& Shapiro, S.~L.\ 1995, \apj, 438, 887 
\bibitem[Seaton et al.(1994)]{seaton94} Seaton, M.~J., Yan, Y., Mihalas, D., \& Pradhan, A.~K.\ 1994, \mnras, 266, 805 
\bibitem[Semenov et al.(2003)]{semenov03} Semenov, D., Henning, T., Helling, C., Ilgner, M., \& Sedlmayr, E.\ 2003, \aap, 410, 611 
\bibitem[Shen et al.(2016)]{shen16} Shen, R.-F., Nakar, E., \& Piran, T.\ 2016, \mnras,  
\bibitem[Shu et al.(1979)]{shu79} Shu, F.~H., Anderson, L., \& Lubow, S.~H.\ 1979, \apj, 229, 223 
\bibitem[Smith et al.(2016)]{smith16} Smith, N., Andrews, J.~E., Van Dyk, S.~D., et al.\ 2016, \mnras, 458, 950
\bibitem[Sobotka et al.(2002)]{sobotka02}Sobotka, P., \v{S}melcer, L., Pejcha, O., et al.\ 2002, Information Bulletin on Variable Stars, 5336, 1 
\bibitem[Soker \& Tylenda(2003)]{soker03} Soker, N., \& Tylenda, R.\ 2003, \apjl, 582, L105 
\bibitem[Soker \& Tylenda(2006)]{soker06} Soker, N., \& Tylenda, R.\ 2006, \mnras, 373, 733 
\bibitem[Stamatellos et al.(2007)]{stamatellos07} Stamatellos, D., Whitworth, A.~P., Bisbas, T., \& Goodwin, S.\ 2007, \aap, 475, 37 
\bibitem[Sytov et al.(2007)]{sytov07} Sytov, A.~Y., Kaigorodov, P.~V., Bisikalo, D.~V., Kuznetsov, O.~A., \& Boyarchuk, A.~A.\ 2007, Astronomy Reports, 51, 836 
\bibitem[Sytov et al.(2009)]{sytov09} Sytov, A.~Y., Bisikalo, D.~V., Kaigorodov, P.~V., \& Boyarchuk, A.~A.\ 2009, Astronomy Reports, 53, 428 
\bibitem[Thompson(2011)]{thompson11} Thompson, T.~A.\ 2011, \apj, 741, 82
\bibitem[Tomida et al.(2013)]{tomida13} Tomida, K., Tomisaka, K., Matsumoto, T., et al.\ 2013, \apj, 763, 6 
\bibitem[Tomida et al.(2015)]{tomida15} Tomida, K., Okuzumi, S., \& Machida, M.~N.\ 2015, \apj, 801, 117
\bibitem[Tout et al.(1989)]{tout89} Tout, C.~A., Eggleton, P.~P., Fabian, A.~C., \& Pringle, J.~E.\ 1989, \mnras, 238, 427 
\bibitem[Tylenda(2005)]{tylenda05} Tylenda, R.\ 2005, \aap, 436, 1009 
\bibitem[Tylenda \& Soker(2006)]{tylenda06} Tylenda, R., \& Soker, N.\ 2006, \aap, 451, 223 
\bibitem[Tylenda et al.(2011)]{tylenda11} Tylenda, R., Hajduk, M., Kami{\'n}ski, T., et al.\ 2011, \aap, 528, A114 
\bibitem[Tylenda et al.(2013)]{tylenda13} Tylenda, R., Kami{\'n}ski, T., Udalski, A., et al.\ 2013, \aap, 555, A16 
\end{thebibliography}
\end{document}